\documentclass[paper]{ieice}
\usepackage{graphicx,xcolor}
\usepackage[fleqn]{amsmath}
\usepackage{newtxtext}
\usepackage[varg]{newtxmath}

\field{B}
\title{Deep-Reinforcement-Learning-Based Distributed\\Vehicle Position Controls for Coverage Expansion in mmWave V2X}
\authorlist{%
\breakauthorline{2}
\authorentry{Akihito Taya}{s}{kyoto-u}\MembershipNumber{1102117}
\authorentry[nishio@i.kyoto-u.ac.jp]{Takayuki Nishio}{m}{kyoto-u}\MembershipNumber{1002899}
\authorentry{Masahiro Morikura}{f}{kyoto-u}\MembershipNumber{8110692}
\authorentry{Koji Yamamoto}{e}{kyoto-u}\MembershipNumber{0212512}
}
\affiliate[affiliate label]{The author is with the Graduate School of Informatics, Kyoto University, Yoshida-honmachi, Sakyo-ku, Kyoto, 606--8501 Japan.}
\received{20xx}{1}{1}
\revised{20xx}{1}{1}

\usepackage{bm}
\usepackage{cite}
\usepackage{mathtools}
\usepackage{enumerate}
\usepackage{algorithm}
\usepackage{algorithmicx}
\usepackage{algpseudocode}
\usepackage{xspace}
\usepackage{color}
\usepackage{soul}
\usepackage[most]{tcolorbox}

\renewcommand{\]}{\right]}
\renewcommand{\(}{\left(}
\renewcommand{\)}{\right)}

\newcommand{\rmm}{R_\mathrm{mm}}
\newcommand{\rctrl}{R_\mathrm{c}}  
\newcommand{\Rd}{d}                
\newcommand{\Rex}[1]{r_{#1}}       
\newcommand{\Rvi}{r_\mathrm{V2I}}  
\newcommand{\Rvv}{r_\mathrm{V2V}}  
\newcommand{\Nlane}{N_\mathrm{l}}  

\newcommand{\Cex}[1]{C_{#1}}           
\newcommand{\Cexavg}{C_\mathrm{avg}}   

\newcommand{\Nall}{n}                  
\newcommand{\Nctrl}{n_\mathrm{c}}      
\newcommand{\Nmm}{n_\mathrm{m}}        

\newcommand{\Vall}{P_\mathrm{a}}
\newcommand{\Vctrl}{P_\mathrm{c}}            
\newcommand{\Vunc}{P_\mathrm{u}}
\newcommand{\Vmicro}{P_\mathrm{n}}
\newcommand{\Vcspace}{\mathcal{P}}

\newcommand{\ActSpace}{\mathcal{A}}
\newcommand{\ValueFunc}{V}
\newcommand{\Adv}{A}
\newcommand{\si}{s}
\newcommand{\acti}{a}
\newcommand{\ri}{r}
\newcommand{\thetav}{\theta_\mathrm{v}}
\newcommand{\thetavi}{\theta_{\mathrm{v},i}}
\newcommand{\penal}{r_\mathrm{p}}
\newcommand{\range}{\mathcal{R}}
\newcommand{\tmax}{t_\mathrm{max}}
\newcommand{\vfcoef}{c_\mathrm{v}}
\def\PT{PT\xspace}
\def\PTCL{PTCL\xspace}
\def\PTDL{PTDL\xspace}

\def\figsizeIdea{0.46\textwidth}
\def\figsizeSM{0.47\textwidth}
\def\figsizeAgent{0.43\textwidth}
\def\figsizeDNN{0.43\textwidth}
\def\figsizeData{0.42\textwidth}

\begin{document}
\maketitle
\begin{summary}
In millimeter wave (mmWave) vehicular communications,
multi-hop relay disconnection by line-of-sight (LOS) blockage is a critical problem,
particularly in the early diffusion phase of mmWave-available vehicles,
where not all vehicles have mmWave communication devices.
This paper proposes a distributed position control method
to establish long relay paths through road side units (RSUs).
This is realized by a scheme via which
autonomous vehicles change their relative positions
to communicate with each other via LOS paths.
Even though vehicles with the proposed method
do not use all the information of the environment
and do not cooperate with each other,
they can decide their action (e.g., lane change and overtaking)
and form long relays only using information of their surroundings
(e.g., surrounding vehicle positions).
The decision-making problem is formulated as a Markov decision process
such that autonomous vehicles can learn a practical movement strategy
for making long relays by a reinforcement learning (RL) algorithm.
This paper designs a learning algorithm based on
a sophisticated deep reinforcement learning algorithm,
asynchronous advantage actor-critic (A3C),
which enables vehicles to learn a complex movement strategy quickly
through its deep-neural-network architecture and multi-agent-learning mechanism.
Once the strategy is well trained,
vehicles can move independently to
establish long relays and connect to the RSUs via the relays.
Simulation results confirm that
the proposed method can increase the relay length and coverage
even if the traffic conditions and penetration ratio of mmWave communication devices
in the learning and operation phases are different.
\end{summary}

\section{Introduction}
Autonomous and connected vehicles can not only achieve
safe and efficient transportation
but can also provide several intelligent services
such as real-time detailed maps, vehicular cloud computing,
cooperative perception, and infotainment
\cite{conectedvehicles,IoV,MmWaveVanetSurvey,VCC}.
Some of these services require
high-throughput vehicle-to-infrastructure (V2I) communication,
which is one of the reasons why
millimeter-wave (mmWave) vehicular communications have attracted considerable attention
\cite{MmWaveVanetSurvey,eband,PathLossPrediction,wu2017cooperative}.

Although mmWave communications can achieve high-throughput data transmission,
their communication ranges tend to be shorter than microwave communications
because of their high attenuation.
This makes the communication coverage of road side units (RSUs) smaller.
For microwave vehicular communications,
multi-hop relaying has been studied
to extend the coverage of RSUs \cite{ChainCluster,DynClusterBase,salvo2012road}.
Multi-hop relaying is significantly more important for mmWave communications
because of their short communication range.
However, mmWave relaying presents a major challenge in the form of its disconnection problem.
In mmWave vehicle-to-vehicle (V2V) relaying,
the line-of-sight (LOS) path is easily blocked by other vehicles,
particularly vehicles that do not have the capability of mmWave relaying.
Therefore, mmWave V2V relay networks can be disconnected and fragmented frequently
because mmWave signals are severely attenuated by the blockages.
This disconnection induced by the blockage is critical,
particularly in the diffusion phase of mmWave-available vehicles,
where not all vehicles have mmWave communication devices.

\begin{figure}[t] \centering
  \includegraphics[width=\figsizeIdea]{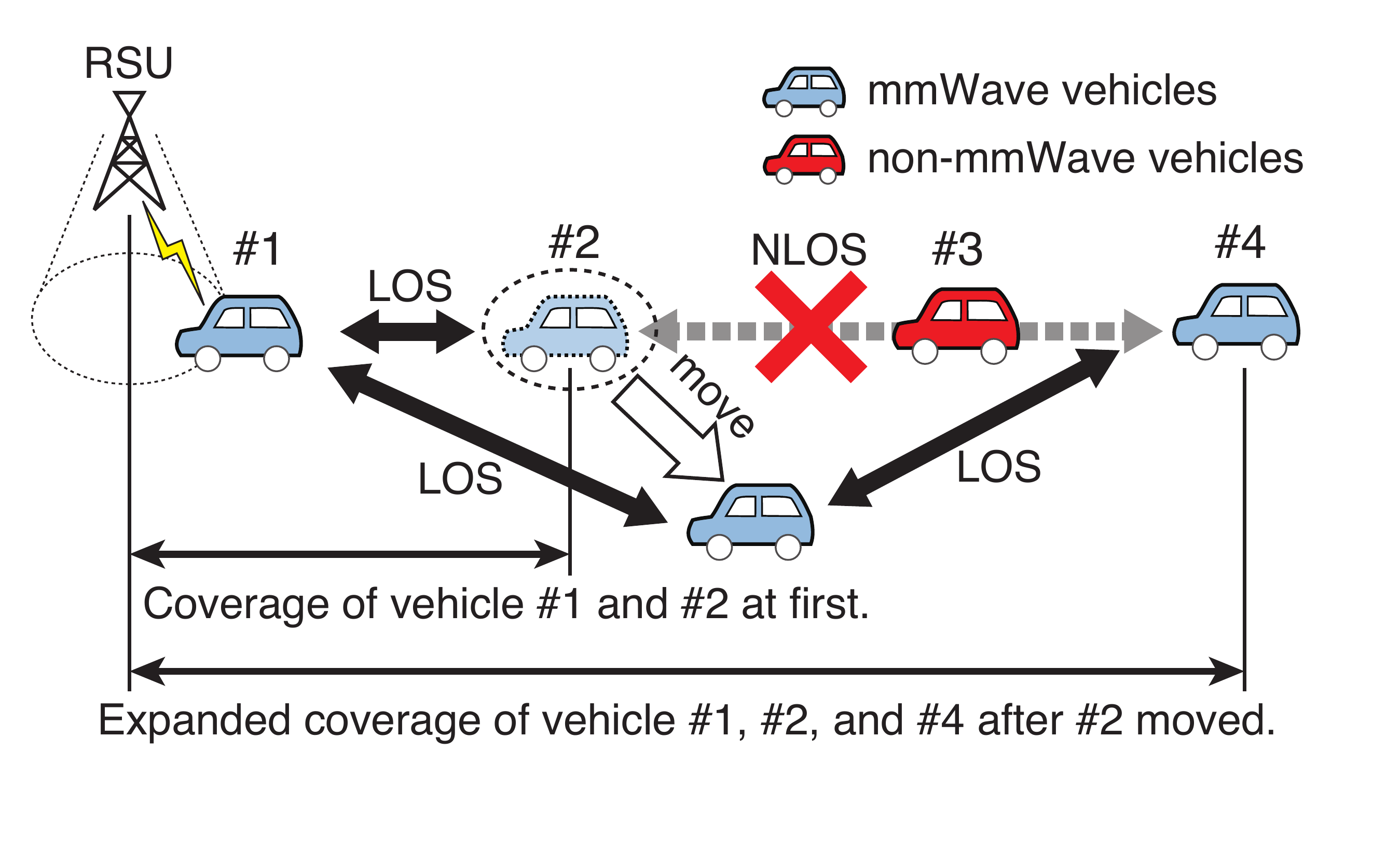}
  \caption{Vehicle \#3 initially blocks the path between vehicles \#2 and \#4.
          When vehicle \#2 changes its position,
          vehicles \#2 and \#4 communicate with each other via a LOS path;
          thus, the multi-hop relay becomes long.}
  \label{fig:proposemethod}
\end{figure}

This paper studies a vehicle position control method
to solve the relay disconnection problem
and extend the coverage of RSUs.
The proposed method enables position-controllable vehicles, such as autonomous vehicles,
to change their relative positions,
where LOS paths are available and V2V relays can be connected
by lane change or overtaking.
Figure~\ref{fig:proposemethod} shows a simple example of our method.
Let vehicles \#1, \#2, and \#4 have mmWave communication devices,
while vehicle \#3 does not have one.
Vehicles \#1 and \#2 are initially connected with each other on mmWave channels,
but vehicles \#2 and \#4 are not connected because of the blockage by vehicle \#3.
In this case,
by moving vehicle \#2 to a position
where a LOS path between vehicle \#2 and \#4 becomes available,
the relay length can be extended and thus,
the RSU coverage is extended to include vehicle \#4.

Cooperative position controls of vehicles have been studied
in \cite{platoonManagement,platoonIVC,infoFlow,coFormation}.
While they focus on improving the stability and robustness of formation and platooning control,
they do not consider improving communication quality.
In the fields of robots and wireless sensors,
movement or position control methods
to improve the connectivity of multi-hop communications have been discussed
\cite{MovementCtrl,vehicularVFA}.
However, these prior works do not focus on the relay length,
and they consider conventional microwave communications where no harmful blockage occurs.

Our previous work proposed a vehicle position control method
for coverage expansion in mmWave vehicular networks
and revealed the maximum gain of coverage improvement
by optimizing vehicle positions \cite{taya2018coverage}.
However, this work uses a centralized algorithm,
which is not sufficiently scalable,
and does not consider the movement of each vehicle to its optimal position.
To maximize the relay length,
both positions of vehicles and their movement paths should be optimized.
However, the optimization problem of collision-free path planning of multiple vehicles
is an NP-hard problem \cite{raja2012optimal}.
Therefore, the optimal solution cannot be obtained
even if centralized manners are applied.
Moreover, centralized algorithm is not practical for vehicle controls
because it cannot be applied when lots of vehicles exist in wide area,
thus a distributed mechanism is required
that enables vehicles to expand relay length
by moving to good positions without collision.

In this paper, we develop a deep-reinforcement-learning (DeepRL)-based vehicle position control method,
where each vehicle distributedly decides its movement
without knowledge of the optimal position
or global information other than that of the surrounding vehicles.
Reinforcement learning (RL) is introduced to obtain a movement strategy
via which vehicles decide their movement
in order to increase coverage based on their own information,
such as locations of the surrounding vehicles.
Because such a strategy is high-dimensional and non-linear function,
we employ DeepRL, which combines RL algorithms and deep learning methods.
To increase the learning speed by updating the strategy cooperatively,
we leverage a distributed DeepRL algorithm,
asynchronous advantage actor-critic (A3C) \cite{a3c},
which was originally developed for parallel processing.
In A3C, multiple agents update the shared strategy cooperatively,
while acting in multiple environments independently.
By applying the cooperative-strategy-updating scheme to our coverage-improvement problem,
multiple vehicles can cooperatively and efficiently learn the strategies;
this allows them to make decisions without communicating with each other.
Because DeepRL algorithms achieve high performance
in problems where the environment state is represented as a pictorial image,
we design a three-dimensional state representation like color images
for the coverage expansion problem.

Our simulation results show that
vehicles can learn the coverage-increasing strategy
using the DeepRL algorithm.
We also demonstrate that the learned strategies achieve high performance
even when the traffic conditions and the penetration ratio of mmWave-available vehicles
are changed from that in the learning phases.

The contributions of this paper are summarized as follows:
\begin{enumerate}[(1)]
\item We present a solution for blockage problems in mmWave vehicular networks.
The proposed method is based on position control of vehicles,
whereas conventional works \cite{ChainCluster,DynClusterBase,salvo2012road}
have assumed that vehicle positions are specified and not controllable.
\item We present a distributed DeepRL-based vehicle movement-control algorithm.
By leveraging a DeepRL algorithm, A3C,
our algorithm enables vehicles to learn a movement strategy,
which involves complex mapping from surrounding vehicle positions to movement action.
\item The simulation results justify that the proposed algorithm achieves high performance
even when the environment conditions
(e.g., the penetration ratio of mmWave-communicable vehicles and vehicle density)
of learning and test phases are not the same.
\end{enumerate}

The rest of this paper is organized as follows:
Related works are discussed in Section \ref{sec:relatedworks}.
Then, our system model is described in Section \ref{sec:system},
and the DeepRL-based approach
to improve coverage is presented in Section \ref{sec:rl}.
Finally, simulation results are presented in Section \ref{sec:results}
and the conclusions are drawn in Section \ref{sec:conclusion}.

\section{Related Works} \label{sec:relatedworks}
\subsection{Vehicular Networks}
One of the current available vehicular communication protocols is  
dedicated short range communication (DSRC).
DSRC, which has been standardized as IEEE 802.11p,
uses the 5.8--5.9\,GHz band \cite{jiang2006design}.
Another protocol is cellular vehicle-to-everything (C-V2X),
which is specified in the third generation partnership project (3GPP) Release 14 \cite{molina2017lte}.
Unfortunately, these protocols cannot meet
the increasing demand of high-data-rate vehicular communications
for sharing enormous sensor data and providing infotainment
because of their limited bandwidth.

Therefore, mmWave communication is an essential technique
for vehicular communications to meet the increasing demand.
MmWave communications offer huge bandwidth (e.g., 9\,GHz in 60\,GHz band)
and realize beyond-Gbit/s throughput.
In addition, offloading data transmission of non-safety applications
from the microwave bands to the mmWave bands
can reduce the pressure on the microwave band,
and should be used for critical applications
such as vehicle collision warning systems and self-driving systems.
The authors of \cite{MmWaveVanetSurvey} provide
a survey of mmWave vehicular communications,
including detailed analysis of
mmWave spectrum, PHY, and MAC designs for mmWave communications
that can be applied to vehicular communications.
MAC protocols for mmWave multi-hop relaying for vehicular networks
have been studied in \cite{raloha,wu2017cooperative}.
In \cite{raloha}, an ALOHA-based protocol is evaluated for a one-lane highway scenario,
and it is demonstrated that multi-hop achieves better performance than single-hop
in disseminating information to a certain number of vehicles.
The authors of \cite{wu2017cooperative} study content delivery using heterogeneous networks
consisting of the licensed Sub-6\,GHz band, DSRC, and mmWave communications.
They introduce fuzzy logic to select efficient cluster heads
considering the vehicle velocity, vehicle distribution, and antenna height.
Whereas these studies assume that vehicle movements are given,
we propose to control vehicle movements in order to avoid blockage.

Multi-hop relaying for increasing the coverages of RSUs
has been studied for the microwave vehicular adhoc network (VANET).
The authors of \cite{ChainCluster}
propose multi-hop V2V relaying
to compensate for the sparsity of RSUs.
\cite{DynClusterBase} minimizes the number of vehicles
that communicate with RSUs by constructing clusters
where messages are transmitted via V2V relays.
The authors of \cite{salvo2012road} propose
a forwarding algorithm to extend the coverage of RSUs for urban areas.
The algorithm realizes multi-directional dissemination in intersections
by considering the geographic positions of vehicles.
Although these prior studies show high performance when the vehicle density is high,
relay disconnections by obstacles are not considered
because these studies assume microwave VANETs.
We leverage the mobility controllability of autonomous vehicles
to connect vehicles via LOS paths.
Such a topology-modifying scheme is one of the differences
between our method and conventional methods.

Position control of vehicles has been studied
in the literature of cooperative vehicle platoon and formation controls
\cite{platoonManagement,platoonIVC,infoFlow,coFormation}.
The authors of \cite{platoonManagement} develop a platoon-management protocol based on VANET.
In the protocol,
the platoon leader sends beacons that contain platoon parameters
to followers in a multi-hop manner,
and each follower maintains an appropriate inter-vehicle space.
\cite{platoonIVC} proposes
an intra-platoon management strategy
that ensures platoon stability under the presence of communication delays.
The authors of \cite{infoFlow,coFormation}
study formation controls for cooperative vehicles.
They utilize graph theory and
demonstrate the robustness of their methods.
In \cite{UAVformation},
a formation-control method based on potential fields
is proposed for unmanned aerial vehicles.
Although these methods, which rely on vehicular communication techniques,
improve the stability or robustness of vehicle formations,
they are not intended to improve communication quality.
Moreover, they assume that the desired patterns
or inter-vehicle distances are given,
whereas our work focuses on
finding the relative positions
that increase the coverage of wireless networks.
To our knowledge, such approaches
that control vehicle positions on the road
in order to improve communication quality
have not been studied.

Some researchers of robot control systems, not vehicles,
have proposed movement control or positioning schemes
to improve the connectivity of multi-hop networks.
The authors of \cite{MovementCtrl} propose a block-movement algorithm
to construct fault-tolerant adhoc networks for autonomous multi-robot systems.
However, coverage expansion is not discussed in \cite{MovementCtrl}.
As a decentralized deployment method,
virtual force algorithms (VFAs)
are introduced in \cite{NovelVF,NovelDeploymentVF}.
These algorithms focus on increasing the sensor coverage of wireless sensor networks (WSNs)
under the constraints of the desired connectivity.
VFAs are also used in \cite{vehicularVFA}
for vehicle self-deployment in order to form fault-tolerant adhoc networks.
The authors of \cite{vehicularVFA}
utilize an evolutionary algorithm (EA)
to determine the velocity
and virtual forces for fitness values of the EA.
VFAs perform well in these studies
because they assume microwave communications,
in which connection quality between two nodes
increases as they become closer.
In mmWave communications, however,
blockage effects decrease the connectivity
of non-line-of-sight (NLOS) communications;
thus, mmWave connectivity, as a function of positions, has local maxima.
Therefore, VFAs, which are based on gradients,
are not expected to improve mmWave connectivity
because vehicles are trapped at the local maxima.

Note that other approaches to solve the blockage problem are discussed,
and beamforming is a promising technique
that controls the directionality of array antennas
to leverage a strong reflected signal \cite{mmWaveBeamForming, mmWaveBFindoor, radaraidedV2I}.
Our work does not conflict with beamforming
and can be used simultaneously.
For example, when a LOS path is not available but a NLOS path with a strong signal is available,
beamforming solves the blockage problem.
However, if there are no LOS and strong NLOS paths,
our method moves vehicles to leverage a strong path and extend the relay length.

\subsection{Reinforcement Learning}
Reinforcement learning has been developed
to solve the problem of a mapping from situations to actions
so as to maximize a reward in an environment \cite{RLintro,nguyen2017system}.
In RL, an agent learns the mapping by performing actions
and observing the results of the actions in the environment.
Although table-based Q-learning,
which is one of the most widely used RL algorithms,
is guaranteed to solve the simple Markov decision process (MDP),
it cannot be applied to large-scale problems
due to a limitation of computational memory capacity.
Because of recent studies,
RL using deep neural networks (DNN) for function approximation
can solve large-scale problems
such as video games \cite{dqn}
and the ancient board game Go \cite{alphago}.
Moreover, multi-process learning algorithms,
such as general reinforcement learning architecture (Gorila) \cite{gorila} and A3C \cite{a3c},
have been developed to improve the convergence efficiency and learning speed.
In A3C, multiple agents learn their strategies in independent environments
and asynchronously share the strategies.
We utilize A3C to solve our coverage-improvement problems
in order to improve learning speed by cooperative learning.

\begin{figure}[t] \centering
  \includegraphics[width=\figsizeSM]{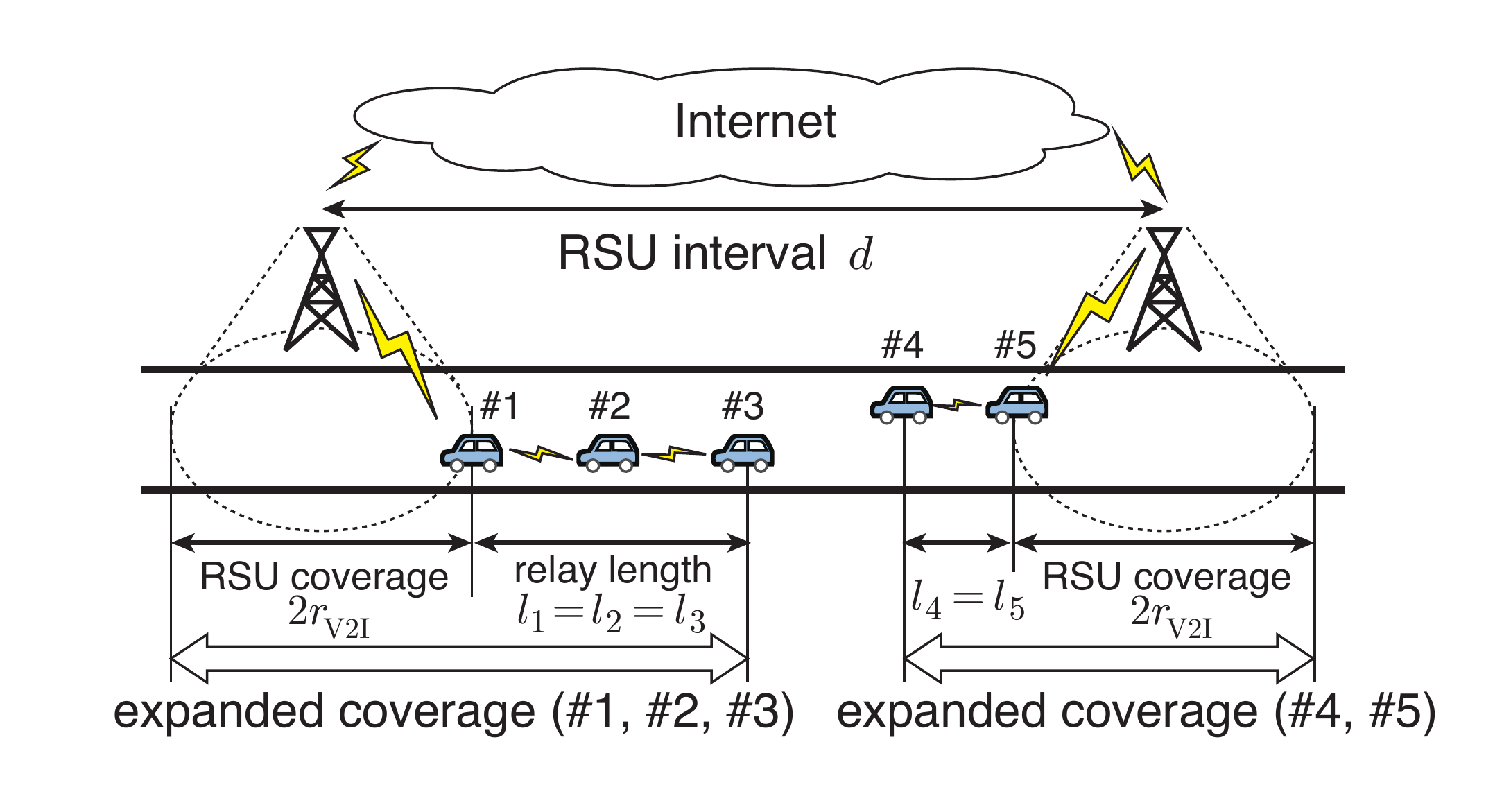}
  \caption{Coverage is expanded through multi-hop relaying.
          $l_i$ denotes the length along the $x$ axis
          of the relay chains to which vehicle $i$ belongs.
          The expanded mmWave V2I communication range of vehicle $i$ can be expressed as
          $2\Rvi + l_i$.
  }
  \label{fig:systemmodel}
\end{figure}
\section{System Model} \label{sec:system}
Figure~\ref{fig:systemmodel} illustrates our system model.
We consider a straight $\Nlane$-lane highway
with mmWave RSUs deployed at intervals of $\Rd$.
We only consider a single direction
as our aim is to create relay chains
that remain unchanged for a long time.
For simplicity, we assume that vehicles are located at grid points.
We assume that the average velocities of all the vehicles are nearly the same,
which is confirmed in \cite{accAutomata},
where the authors demonstrate that
velocity variance decreases
as the ratio of vehicles with adaptive cruise control increases,
which is one of the key technologies in autonomous vehicles.
We also consider two cases in the simulation evaluations:
one where the velocity is constant
and the other where it fluctuate.
When the vehicles move at a constant and equal velocity,
the relative positions of the vehicles do not change.
When the velocity fluctuates,
the movements on the relative positions can be modeled as random walks.
A region of interest (RoI) also moves with the vehicles
at the same velocity as them.

A simplified mmWave communication model is assumed
in which vehicles can be connected
if and only if the distance between the vehicles is less than a certain distance
and a LOS path is available.
We also assume that link qualities of the mmWave channels are ideally predictable.
This assumption is not unrealistic considering state-of-the-art prediction methods.
For example, the authors of \cite{PathLossPrediction,AnalysisUrbanMmWave}
present a mmWave propagation loss model for V2V communications.
Another approach is proposed in \cite{MLbasedTrhoughputPred,RNNbasedRSSPred},
where the authors predict the throughput of mmWave communications
and received signal strength of mmWave radios using image sensor data.

There are non-mmWave vehicles and mmWave vehicles,
which are subdivided into controllable mmWave vehicles
and uncontrollable mmWave vehicles.
Whereas the non-mmWave vehicles only use microwave communication systems,
mmWave vehicles use both microwave and mmWave communication systems
and constitute the multi-hop relay chains for increasing coverage.
Microwave communication systems, such as DSRC and C-V2X,
enable all the vehicles to share
information on vehicle positions and their control signals,
whose packet sizes are very small.
We assume that the effect of transmission loss and delay is negligible
because the duration for a transmission is less than 100\,ms
which is much less than the decision-making interval of vehicles.
For example, in DSRC, the vehicles broadcast their position every 100\,ms \cite{jiang2006design}.
Therefore, the vehicles can retransmit dropped packets several times
and the probability that the vehicles fail to receive all the retransmitted packets is very low.
The mmWave vehicles can access RSUs over mmWave channels
within a distance of $\Rvi$
and communicate with each other only when they are within a distance of $\Rvv$
and there are no vehicles blocking their LOS paths.
The controllable vehicles can change their relative positions on the road
in order to expand coverage,
while
uncontrollable vehicles do not change their relative positions or move randomly
because they have other driving strategies or are driven by humans.

Let $\Nall$, $\Nmm$, and $\Nctrl$ denote
the total number of vehicles, all mmWave vehicles,
and the number of controllable mmWave vehicles in the RoI, respectively.
$\rmm\coloneqq\Nmm/\Nall$ and $\rctrl\coloneqq\Nctrl/\Nmm$ denote
the ratio of the number of the mmWave vehicles
to the total number of vehicles
and the ratio of the number of the controllable vehicles
to the total number of mmWave vehicles, respectively.
Let $\Vall$, $\Vmicro$, $\Vctrl$, and $\Vunc$ denote
the sets of all grid positions in the RoI,
the non-mmWave vehicle positions,
the controllable mmWave vehicle positions,
and the uncontrollable mmWave vehicle positions, respectively.
$\Vctrl \in \Vcspace$ is a variable
that the proposed method controls,
where $\Vcspace \coloneqq 2^{\Vall \setminus (\Vmicro \cup \Vunc)}$
represents a power set of $\Vall \setminus (\Vmicro \cup \Vunc)$.

We define a quality metric for the proposed method
as the proportion of areas where vehicles can connect to RSUs
through multi-hop relay chains
based on the coverage of each vehicle.
When using mmWave vehicles as relay nodes,
the mmWave V2I communication range of vehicle $i$ is expressed as
$\Rex{i} = 2\Rvi + l_i(\Vctrl),$
as shown in Fig.~\ref{fig:systemmodel}.
Here, $l_i(\Vctrl)$ denotes the length along the $x$ axis of the relay chains to which vehicle $i$ belongs.
Because RSUs are deployed at intervals of $\Rd$,
it is sufficient to consider $\Rd$-long sections of road.
The coverage for vehicle $i$ is expressed as follows:
\begin{align}
  \Cex{i}(\Vctrl) = \min\left\{\frac{\Rex{i}}{\Rd}, 1\right\}. \label{eq:cexratio}
\end{align}
Note that the coverage $\Cex{i}(\Vctrl)$ increases linearly
with the length of the multi-hop relay chain.
Now, we define the average of the vehicles' coverage as a quality metric, as follows:
\begin{align}
  \Cexavg(\Vctrl) \coloneqq \frac{1}{\Nmm} \sum^{\Nmm}_{i = 1} \Cex{i}(\Vctrl).
\end{align}

\begin{figure}[t] \centering
  \includegraphics[width=\figsizeSM]{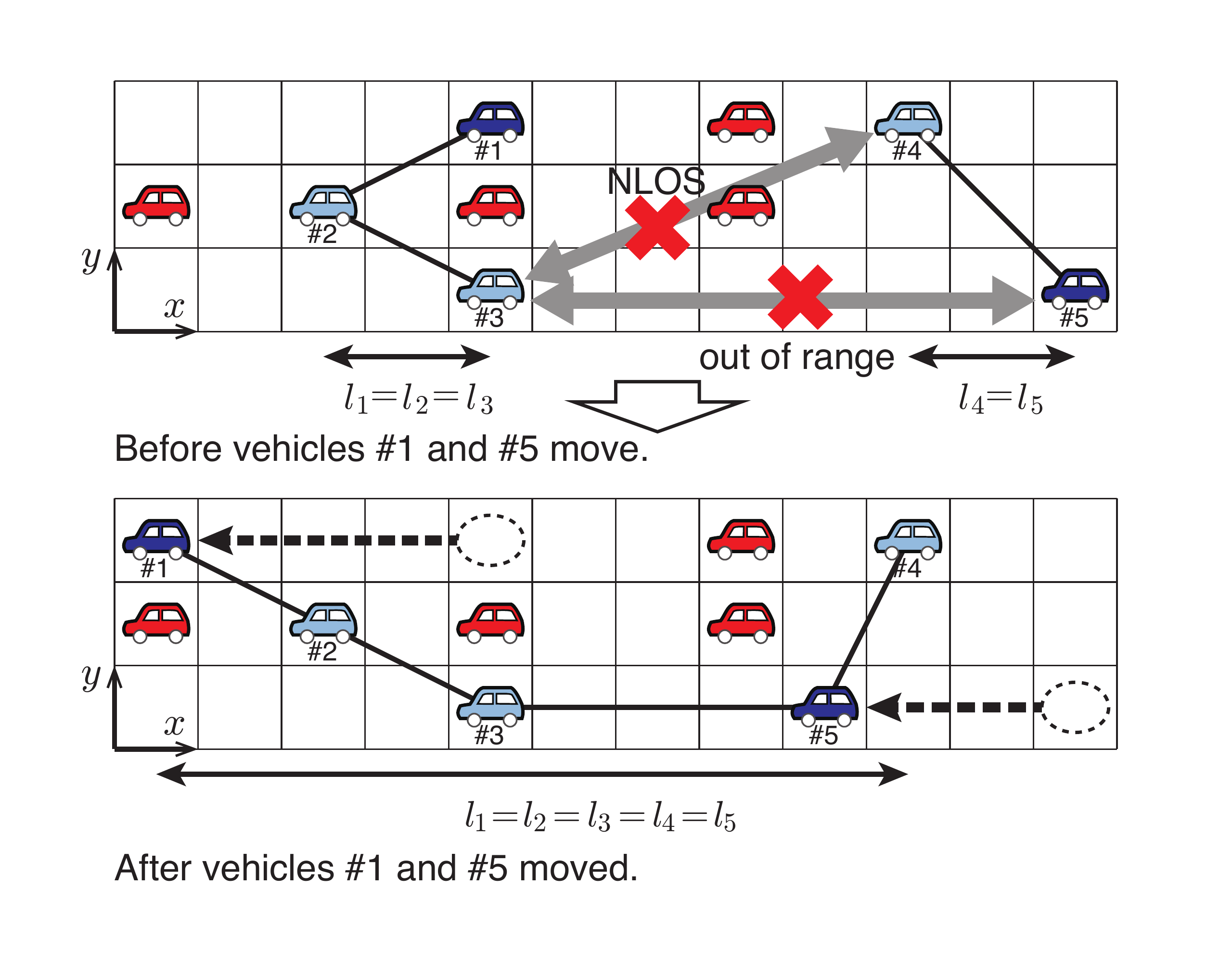}
  \caption{
    Non-mmWave, controllable mmWave, and uncontrollable mmWave vehicles
    are depicted as
    red, dark blue, and light blue vehicles, respectively.
    The mmWave relay length is extended by changing the positions of vehicles \#1 and \#5.
    }
  \label{fig:expandcov}
\end{figure}

Figure~\ref{fig:expandcov} shows
an example of expanding coverage by changing vehicle positions.
Non-mmWave, controllable mmWave, and uncontrollable mmWave vehicles
are depicted as
red, dark blue, and light blue vehicles in Fig.~\ref{fig:expandcov}, respectively.
Vehicles \#1--\#5 are all mmWave vehicles,
but only vehicles \#1 and \#5 are controllable.
Before vehicles \#1 and \#5 move,
the relay chains on the left
and right sides cannot be connected using the mmWave bands
because they are out of communication range or non-mmWave vehicles are blocking the LOS.
However, after vehicles \#1 and \#5 move, vehicles \#3 and \#5 become connected.
At this point, all mmWave vehicles can be connected,
and the length of the relay chain is extended.
$\Cexavg(\Vctrl)$ increases as a result.

\begin{figure}[t] \centering
  \includegraphics[width=\figsizeAgent]{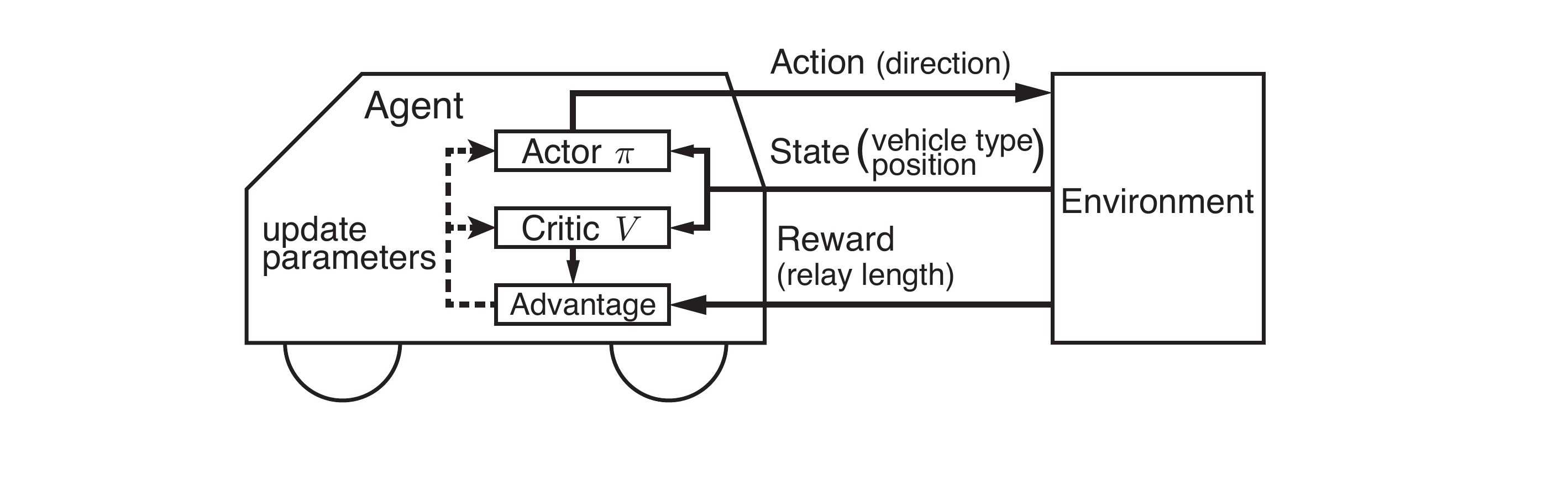}
  \caption{Actor-critic reinforcement learning model}
  \label{fig:a3cmodel}
\end{figure}
\section{Distributed Vehicle Movement Controls} \label{sec:rl}
\subsection{Motivation for Using Reinforcement Learning}
Although we revealed the capability of coverage improvement using vehicle position control in \cite{taya2018coverage},
solving the vehicle-movement problem is still a big challenge.
Because vehicle-moving strategies should be scalable and decentralized for practical usage,
each vehicle should be able to decide its movement for increasing coverage by itself.
Because of conventional microwave communication systems,
information on the surrounding vehicles' positions are available
for each controllable vehicle to decide its movement.
We formulate the decision-making problem as an MDP
such that it can be solved by RL.
Using the RL scheme,
vehicles can learn the optimal positions to increase coverage and go there
without being instructed to.
Because the observed states have large dimensions,
traditional RL methods cannot be applied to our problem;
thus, a DeepRL algorithm is required for function approximation.
We utilize the DeepRL algorithm, A3C,
because it is advantageous in terms of convergence efficiency and learning speed \cite{a3c}.

While vehicles are learning strategies using the A3C algorithm,
they share DNN models with other vehicles.
To this end, we assume there is a model management server
that stores a global model.
The model management server updates the global model
using gradients collected from the vehicles
and distributes the up-to-date model to the vehicles.
These data are transmitted via microwave channels or mmWave relays if they are available.
In this system, vehicles do not always have to be connected to the model management server.
They opportunistically upload their gradient,
and then, the model management server updates the global model.
Once the well-trained model is obtained,
there is no need for vehicles to share DNN models during the operation phase.
Therefore, controllable vehicles decide their movements by using the model
in a decentralized way.
In the following portion of this section,
we explain how vehicles learn movement strategies of coverage expansion using the RL algorithm.
We also describe state designs that improve the performance of increasing coverage.

\subsection{Deep-Reinforcement-Learning-Based Algorithm}
Controllable vehicles, called agents,
interact with an environment
over a number of discrete time steps.
The relationship between an agent and the environment
is shown in Fig.~\ref{fig:a3cmodel}.
At each time step $t$,
agent $i$ observes the surrounding states $\si_{i,t}$
representing other vehicles' positions and types.
The detailed definitions of the states are given by (7), (8), and (9) in Section \ref{sec:state_design}.
The surrounding vehicle information is shared via microwave channels,
and the observation range is limited by the microwave communication range,
which determines the dimension of the states. 
The observation area is limited
by the range of DSRC communications,
where positions and types of vehicles
are broadcasted and shared with each other.
According to an action-selection policy $\pi(\acti_{i,t}|\si_{i,t})$,
agents select an action $\acti_{i,t}$ from a set of actions $\ActSpace$
consisting of \textsl{forward}, \textsl{back}, \textsl{right}, \textsl{left}, and \textsl{stay}.
Note that these actions represent
changing relative positions with respect to the other vehicles.
After moving to the next position,
agent $i$ receives a reward $\ri_{i,t}$.
A reward from each step is defined as:
\begin{align}
  \ri_{i,t} \coloneqq
    \begin{cases*}
      \alpha l_i + \penal  & if an agent selects a prohibited direction \\
      \alpha l_i           & otherwise,
    \end{cases*}
\end{align}
where $l_i$ and $\alpha(>0)$ denote
a relay length and parameter balancing the reward and penalty, respectively.
$\penal(<0)$ denotes a penalty that is added
when an agent selects a direction, where the vehicle should not move.
In order to avoid car accidents,
the prohibited direction is defined as the direction
where other vehicles exist or are moving to
and where the agent gets off the road.
Relay length information is calculated by the following procedure.
Agents broadcast their position information via the relay over the mmWave channel.
Then, the agents calculate the relay length $l_i$
as the length along the road direction between the head and tail vehicles
using shared position information.
Such information sharing can be completed
within a short period compared to the decision-making interval of the RL scheme.

The goal of RL is
to optimize the action-selection policy $\pi(\acti_{i,t}|\si_{i,t})$ 
to maximize the future accumulated reward
$R_{i,t} \coloneqq \sum_{\tau=0}^{\infty} \gamma^\tau \ri_{i,t+\tau}$,
where $\gamma \in [0, 1)$ is a discount factor.
The expectation of $R_{i,t}$ from step $t$ underlying $\pi$
is called the value function,
which is defined as $\ValueFunc^{\pi}(\si_{i,t}) \coloneqq \mathbb{E}\[R_{i,t}|\si_{i,t}\]$.
Each agent acts to maximize its own accumulated reward $R_{i,t}$,
which can be calculated from information obtained by the agent;
thus, our system works in a distributed manner.

We employ A3C \cite{a3c} as the implementation of DeepRL.
A3C learns the optimal policy with sharing models and experiences by multiple agents;
thus, it suits our problem
in which multiple vehicles are required to learn a similar strategy.
Algorithm \ref{alg:a3c} shows the detailed procedure of the A3C learning phase in each episode,
which consists of $T_\mathrm{max}$ time steps.
For more detail, please refer to \cite{a3c}.
A3C parameterizes
the policy function $\pi(\acti_{i,t}|\si_{i,t};\theta)$
and the value function $\ValueFunc(\si_{i,t};\thetav)$,
where $\theta$ and $\thetav$ denote
globally shared parameters of $\pi$ and $\ValueFunc$, respectively.
The globally shared parameters are stored in the model management server.
Agents first downloads the globally shared parameters
and copy them to agent-specific parameters $\theta_i$ and $\thetavi$
when they have started learning (Line~3 in Algo.~\ref{alg:a3c}).
The agents act $\tmax$ times in accordance with their agent-specific policies independently
and store rewards from each step (Lines~4--9 in Algo.~\ref{alg:a3c}).
The agents calculate the gradients of updating parameters $\varDelta\theta_i$ and $\varDelta\thetavi$
at interval $\tmax$ (Lines~10--17 in Algo.~\ref{alg:a3c}).
The calculated gradients are uploaded to the model management server,
and then, the server updates the globally shared parameters $\theta$ and $\thetav$
by gradient ascent and gradient descent, respectively (Lines~18--19 in Algo.~\ref{alg:a3c}).
After sufficient time steps,
the shared policy and value functions each converge to the corresponding optimal functions.
The gradients $\varDelta\theta_i$ and $\varDelta\thetavi$
are calculated as follows:
\begin{align}
  \theta  \colon \varDelta\theta_i
      = & \sum_{\tau=t}^{t+\tmax} \nabla_{\theta_i}
          \log\pi(\acti_{i,\tau}|\si_{i,\tau};\theta_i)
          \Adv(\acti_{i,\tau},\si_{i,\tau};\thetavi) \nonumber \\
        & + \beta \nabla_{\theta_i} H(\pi(\si_{i,\tau};\theta_i)), \label{eq:policygrad} \\
  \thetav \colon \varDelta\thetavi
      = & \vfcoef
          \sum_{\tau=t}^{t+\tmax}
          \partial \(\Adv(\acti_{i,\tau},\si_{i,\tau};\thetavi)\)^2 / \partial \thetavi, \label{eq:valuegrad}
\end{align}
where $\beta$, $H(\pi(\si_{i,\tau};\theta_i))$, and $\vfcoef$
denote a regularization parameter,
the entropy of $\pi(\si_{i,\tau};\theta_i)$,
and a coefficient that balances $\theta$ and $\thetav$, respectively.
$\Adv(\acti_{i,\tau},\si_{i,\tau};\thetavi)$
is an estimation of the advantage
of action $\acti_{i,\tau}$ in state $\si_{i,\tau}$ \cite{a3c},
defined as:
\begin{align}
  \Adv(\acti_{i,\tau},\si_{i,\tau};\thetavi)
      \coloneqq & \sum_{u=0}^{\tau'-1} \gamma^u \ri_{i,\tau+u}
                + \gamma^{\tau'} \ValueFunc(\si_{i,\tau+\tau'};\thetavi) \nonumber \\
                & - \ValueFunc(\si_{i,\tau};\thetavi),
\end{align}
where $\tau'$ is a counter that is incremented at each step in the loop (Line~12--17 in Algo.~\ref{alg:a3c}).

\begin{algorithm}[t!]
  \caption{Algorithm for updating policy and value model parameters.} \label{alg:a3c}
  \begin{algorithmic}[1] 
    \State {Initialize $t \leftarrow 0$.}
    \While {$t < T_\mathrm{max}$}
      \State Download the globally shared parameters and copy them to agent-specific parameters:
             $\theta_i \leftarrow \theta$, $\thetavi \leftarrow \thetav$.
      \State $t_\mathrm{start} \leftarrow t$
      \For {$t \in \{t_\mathrm{start},\dots,t_\mathrm{start}+\tmax \}$}
        \State perform action $\acti_{i,t}$
               according to policy $\pi(\acti_{i,t}|\si_{i,t};\theta_i)$
        \State receive reward $\ri_{i,t}$
        \State $t \leftarrow t + 1$
      \EndFor
      \State Reset gradients $\varDelta\theta_i$ and $\varDelta\thetavi$ to 0.
      \State $R_{i,t} \leftarrow \ValueFunc(\si_{i,t};\thetavi)$
      \For {$\tau \in \{t_\mathrm{start}+\tmax-1,\dots,t_\mathrm{start}\}$}
        \State $R_{i,\tau} \leftarrow \ri_{i,\tau} + \gamma R_{i,\tau+1}$
        \State $\Adv(\acti_{i,\tau},\si_{i,\tau};\thetavi)
                  \leftarrow R_{i,\tau} - \ValueFunc(\si_{i,\tau};\thetavi)$
        \State $\varDelta\theta_i
                  \leftarrow \varDelta\theta_i
                  + \nabla_{\theta_i} \log\pi(\acti_{i,\tau}|\si_{i,\tau};\theta_i)
                    \Adv(\acti_{i,\tau},\si_{i,\tau};\thetavi)
                  + \beta \nabla_{\theta_i} H(\pi(\si_{i,\tau};\theta_i))$
        \State $\varDelta\thetavi
                  \leftarrow \varDelta\thetavi
                  + \vfcoef \partial \(\Adv(\acti_{i,\tau},\si_{i,\tau};\thetavi)\)^2
                  / \partial \thetavi$
      \EndFor
      \State Upload gradients $\varDelta\theta_i$ and $\varDelta\thetavi$ to the server
      \State [SERVER] Update parameters $\theta$ and $\thetav$
    \EndWhile
  \end{algorithmic}
\end{algorithm}

In a common A3C learning phase,
multiple agents in multiple independent environments
share the parameters $\theta$ and $\thetav$ of the learning models.
This model sharing enables efficient convergence
due to the independence between training data sets.
Although there is a single environment in our problem,
the observed states of multiple agents
are different and have little correlation
if their distances are great enough.
DNN models therefore converge due to model sharing.

\subsection{State Designs and Network Models} \label{sec:state_design}
\begin{figure}[t] \centering
  \includegraphics[width=\figsizeDNN]{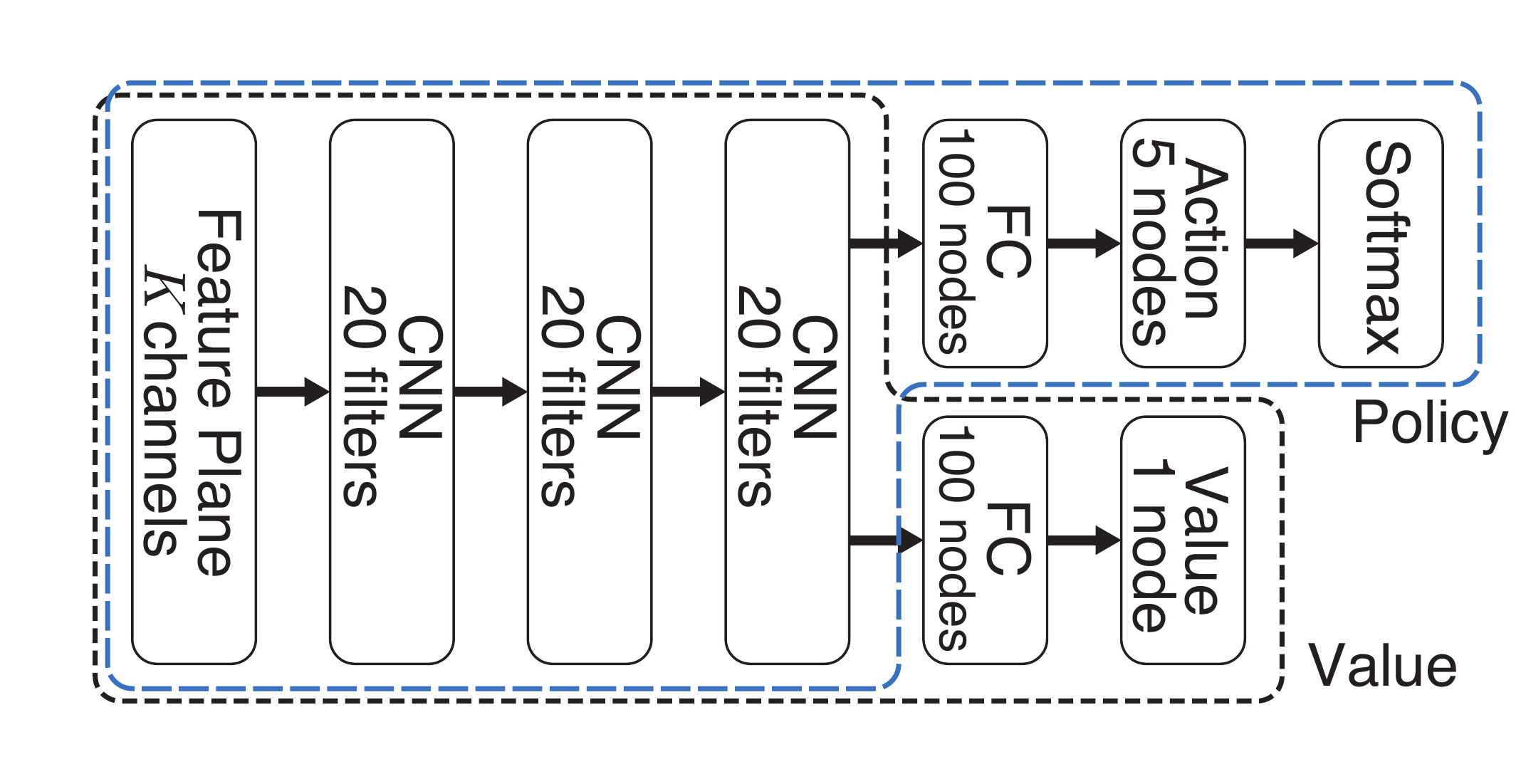}
  \caption{DNN models of policy and value functions.}
  \label{fig:a3c_nn}
\end{figure}

We designed three state definitions:
(1) vehicle positions and types (\PT);
(2) vehicle positions, types, and continuous relay lengths (\PTCL);
and (3) vehicle positions, types, and discrete relay lengths (\PTDL).
We compared the performance of these state definitions in Section \ref{sec:results}
and showed that \PTCL and \PTDL,
which include relay length information,
achieve greater coverage than \PT.

States $\si_{i,t}$ are defined as
three-dimensional $K \times X \times Y$ feature planes,
where $K$, $X$, and $Y$ denote 
the number of features
and the observation ranges of the $x$ and $y$ axes, respectively.
A feature plane design is introduced in AlphaGo \cite{alphago}.
Feature planes represent different features in different planes
and convolutional neural networks (CNN)
are expected to adaptively optimize their parameters to each feature.
Let $\si_{i,t}^{(k,x,y)}$ denote each state element,
where $k$ and $(x,y)$ represent a feature index
and a relative position observed from the vehicle $i$, respectively.
The center of each feature plane represents the position of vehicle $i$.
While $X$ is limited by the microwave communication range,
$Y$ is defined as $Y\coloneqq 2\Nlane - 1$ to cover all lanes.
The state elements $\si_{i,t}^{(k,x,y)}$ in \PT are defined as follows:
\begin{align}
  \si_{i,t}^{(1,x,y)} & \coloneqq
  \begin{cases}
    1 & \text{if a mmWave vehicle positions at } (x,y),\\
    0 & \text{otherwise},
  \end{cases} \label{eq:pt1} \\
  \si_{i,t}^{(2,x,y)} & \coloneqq
  \begin{cases}
    1 & \text{if a non-mmWave vehicle positions at } (x,y),\\
    0 & \text{otherwise},
  \end{cases} \label{eq:pt2} \\
  \si_{i,t}^{(3,x,y)} & \coloneqq
  \begin{cases}
    1 & \text{if } (x,y) \text{ is empty,}\\
    0 & \text{otherwise}.
  \end{cases} \label{eq:pt3}
\end{align}

In order to improve the coverage performance of RL-based methods,
we designed states with additional information
regarding the achievable relay length $\hat{l}_i(x,y)$
when vehicle $i$ would move to position $(x,y)$.
$\hat{l}_i(x,y)$ is calculated by each agent
using vehicle positions and type information.
Each agent is required to estimate whether
a stable mmWave communication link could be established
to calculate $\hat{l}_i(x,y)$.
Link stability can be estimated using path-loss prediction models \cite{PathLossPrediction,AnalysisUrbanMmWave},
which enable vehicles to predict path loss
considering blockage effects caused by other vehicles.
Margins can be considered for the received power required to establish links
in order to avoid misprediction.
Additional \PTCL features are defined as follows:
\begin{align}
  \si_{i,t}^{(4,x,y)}& \coloneqq \rho \hat{l}_i(x,y), \label{eq:ptcl}
\end{align}
where $\rho$ denotes a normalization factor.
The features for $k=1,2,3$ are defined to be the same as \PT.

In contrast to \PTCL,
relay length $\hat{l}_i(x,y)$ of \PTDL
is encoded to one-hot vectors,
which are employed in AlphaGo \cite{alphago}.
State elements for $k\in\{4,\dots,K\}$ are defined as follows:
\begin{align}
  \si_{i,t}^{(k,x,y)}& \coloneqq
  \begin{cases}
    1 & \text{if}\ \hat{l}_i(x,y) \in \range_k, \\
    0 & \text{otherwise},
  \end{cases} \label{eq:ptdl} \\
  \range_4 &\coloneqq \{0\}, \nonumber \\
  \range_{k} &\coloneqq (L_k, L_{k+1}]\ \ \ \text{for}\ k = 5,\dots,K-1, \nonumber \\
  \range_{K} &\coloneqq (L_K,\infty),
\end{align}
where $L_k$ denotes a border series
used to encode relay lengths $\hat{l}_i(x,y)$
into one-hot vectors.

After pre-processing,
states $\si_{i,t}$ are
input into a policy function $\pi$ and a value function $\ValueFunc$.
Their DNN models are shown in Fig.~\ref{fig:a3c_nn}.
The lower CNN layers are shared by two functions,
and the higher fully connected (FC) layers are separated.
Such a shared model is used in \cite{a3c}.

\section{Simulation Evaluations} \label{sec:results}

\subsection{Simulation Setup}
\begin{table}[t] \centering \small \tabcolsep=1mm
  % \caption{Reinforcement learning parameters.}
  \caption{Simulation parameters.}
  \label{tbl:rlparams}
  \begin{tabular}{cc}
    \hline
    Parameters & Values \\
    \hline
    Length of RoI & 1\,km \\
    Number of lane $\Nlane$ & 4 \\
    Number of grids & 200 $\times$ 4 \\
    Lane width & 3.5\,m \\
    RSU interval $\Rd$ & 1\,km \\
    mmWave V2V communication range $\Rvv$ & 50\,m \\
    mmWave V2I communication range $\Rvi$ & 100\,m \\
    Size of the observation area $X \times Y$ & $41 \times 7$ \\
    Normalization factor for \PTCL $\rho$ & $0.005\,\mathrm{m}^{-1}$ \\
    The number of features of \PTDL $K$ & 9 \\
    Boarder series of \PTDL $L_k\ (k=5,\dots,9)$ & $0,25,50,100,150$\\
    Reward balancing parameter $\alpha$ & $0.5\,\mathrm{m}^{-1}$ \\
    Penalty $\penal$ & -2 \\
    Discount factor $\gamma$ & 0.1 \\
    Maximum time steps per episode $T_\mathrm{max}$ & 100 \\
    Number of episodes (learning phase) & 300 \\
    Number of episodes (test phase) & 100 \\
    Update intervals $\tmax$ & 2 \\
    Regularization parameter $\beta$ & 0.01 \\
    Optimizer & Shared RMSProp \cite{a3c} \\
    Coefficient of balancing gradients $\vfcoef$ & 0.5 \\
    \hline
  \end{tabular}
\end{table}

In the simulations,
the vehicles were located at randomly selected grid points.
Let $\lambda$ denote the density of vehicles in each lane.
Vehicle types were determined randomly according to the ratio $\rmm$ and $\rctrl$.
The controllable vehicles had their actions controlled by the proposed RL algorithm.
On the other hand,
the non-mmWave vehicles and uncontrollable vehicles
did not change their relative positions
or selected action among \textit{forward}, \textit{stay}, and \textit{back}
in order to simulate two scenarios
where they drive at equal and constant velocities
and where they change their velocities randomly.
If more than one vehicle selected the same grid,
the vehicle that had the highest priority moved to the selected grid,
while the others did not change their positions.
Penalties were given to
vehicles that could not move to the selected grid.
The priority was defined as follows:
\begin{itemize}
\item Non-mmWave vehicles and uncontrollable vehicles have greater priority than controllable vehicles.
\item Forward vehicles have greater priority if their types are the same.
\item Vehicles with small $y$ position values have greater priority if their types and $x$ position values are the same.
\end{itemize}
The first rule is applied to prevent the proposed method from interrupting the other vehicles' movement.
The second one simulates the realistic driving manners.
To simplify the simulation, we also apply the third one.
The mmWave vehicles could communicate with each other over the mmWave band
if the distance between them was less than 50\,m.
Blocking by other vehicles is detected as follows:
We draw a line from the sender to the receiver
and list the grid sections that cross the line.
If there is no vehicle on the listed grid sections,
the sender and the receiver can communicate with each other via LOS path.
We also assumed that link qualities of the mmWave channels are ideally predictable in our simulations.

We performed python-based simulations
to determine the performance of the movement strategy based on DeepRL.
The RL simulations consist of two phases, the learning and test phases.
In the learning phase,
the controllable vehicles
move around and update their RL models according to Algorithm~\ref{alg:a3c}.
We perform 300 episodes in our simulations.
At the beginning of each episode,
non-mmWave vehicles and uncontrollable vehicles are relocated randomly
in order to obtain a general strategy.
We obtain some models which are learned under different conditions:
the vehicle density, ratio of mmWave vehicles,
and ratio of controllable vehicles.
We assume the conditions are stable while learning each model.
In the test phase,
we evaluate the performance of the globally shared policy models obtained from the learning phases.
At the beginning of the evaluations,
the shared policy models are copied to agent-specific models
and they are not updated while being evaluated.
The conditions in the learning and testing phases can be different,
and we compare the performance of the different models for the same evaluation conditions.
The RL simulation parameters and their values are listed in Table~\ref{tbl:rlparams}.

\begin{figure}[t] \centering
  \includegraphics[width=\figsizeData]{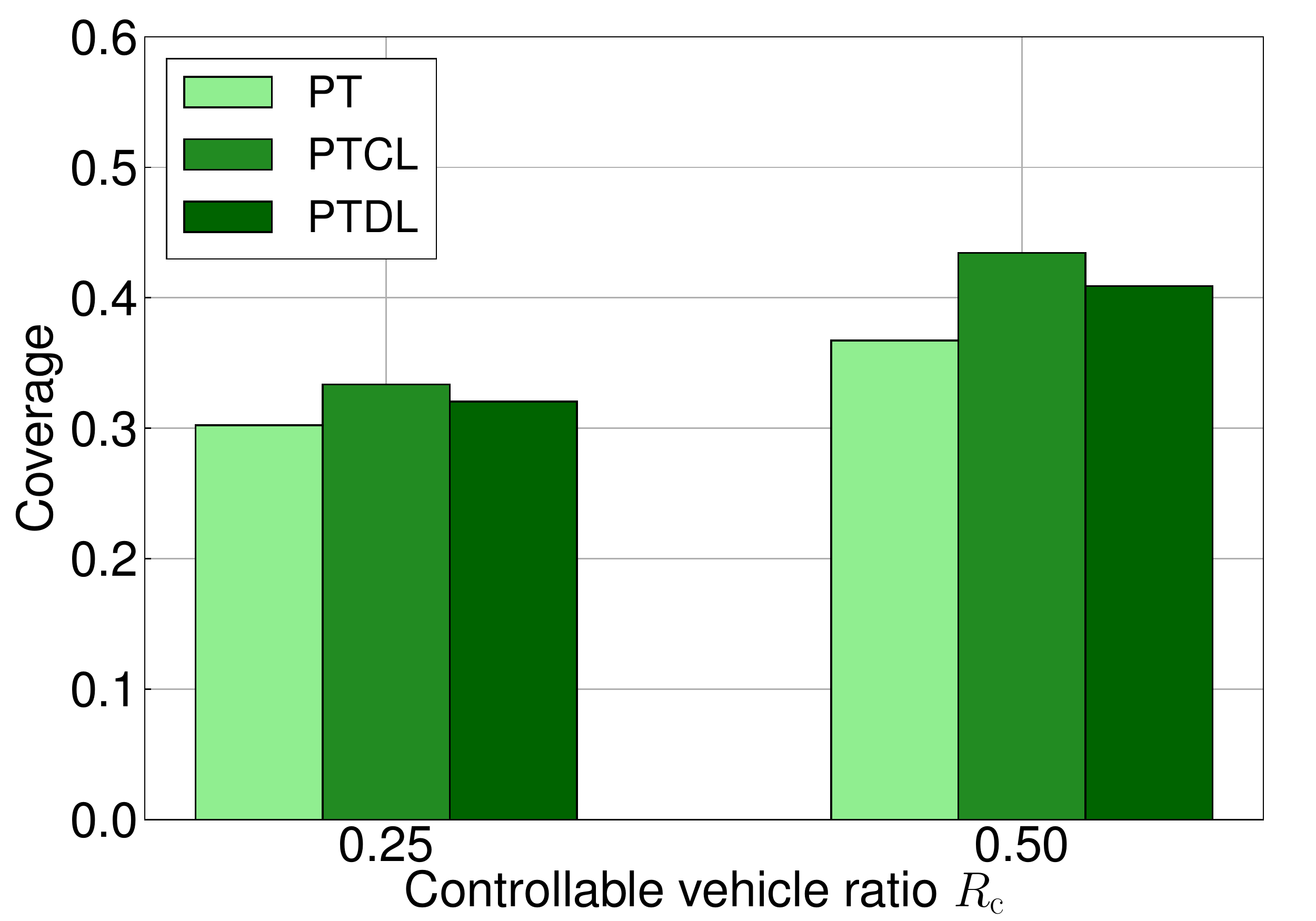}
  \caption{Coverage of different state designs
  when $\lambda=0.02$\,veh/m/lane and $\rmm=0.4$.
  Uncontrollable vehicles move at constant velocities.
  The results of \PTCL and \PTDL are greater coverage than that of \PT.}
  \label{fig:rlstate}
\end{figure}

\begin{figure}[t] \centering
  \includegraphics[width=\figsizeData]{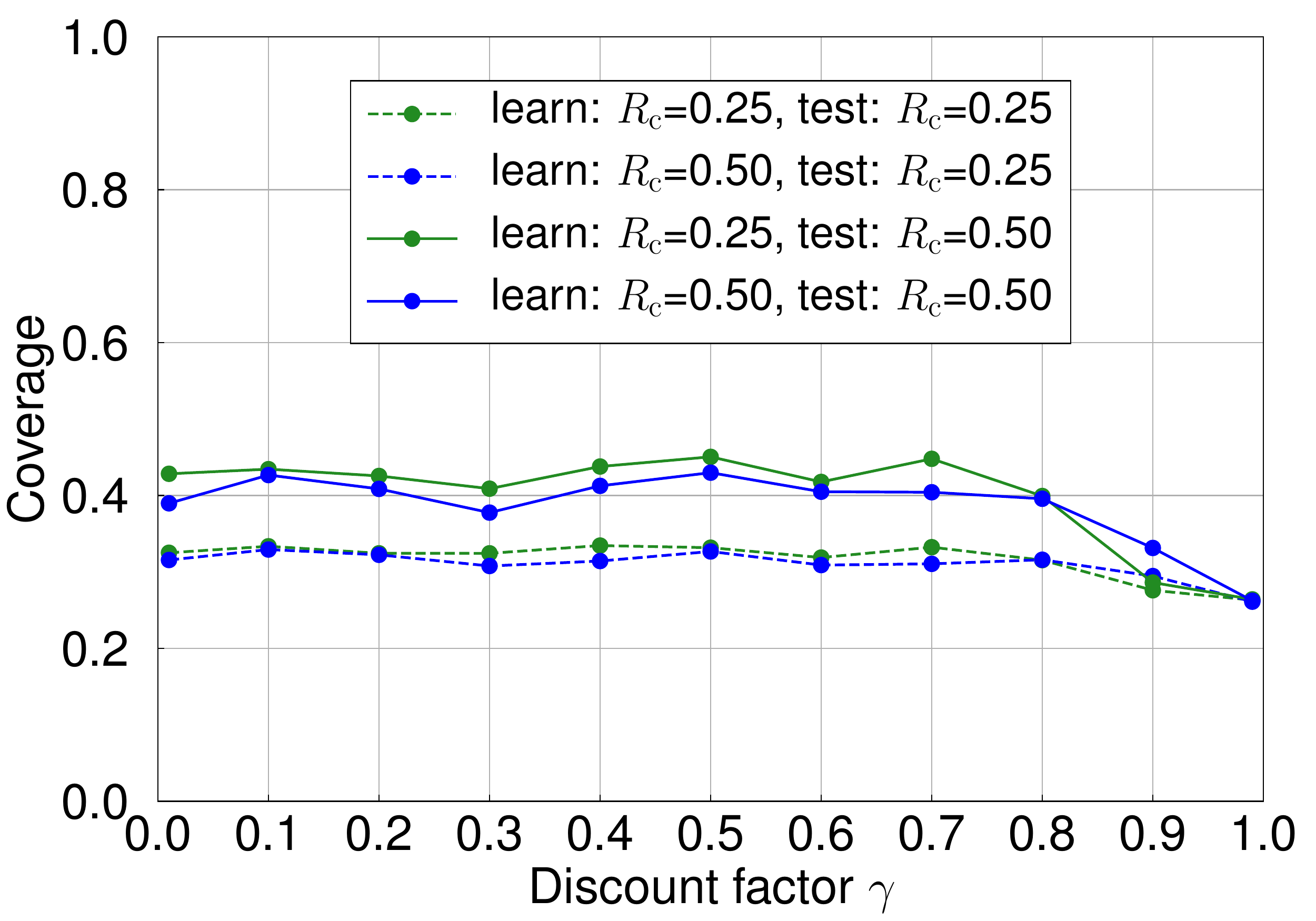}
  \caption{Coverage for different discount factors $\gamma$
  when $\lambda=0.02$\,veh/m/lane, $\rmm=0.4$, and \PTCL is used as the state type.
  Uncontrollable vehicles move at constant velocities.
  Policies learned when $\gamma \le 0.8$ show higher performance.}
  \label{fig:discount}
\end{figure}

\begin{figure}[t] \centering
  \includegraphics[width=\figsizeData]{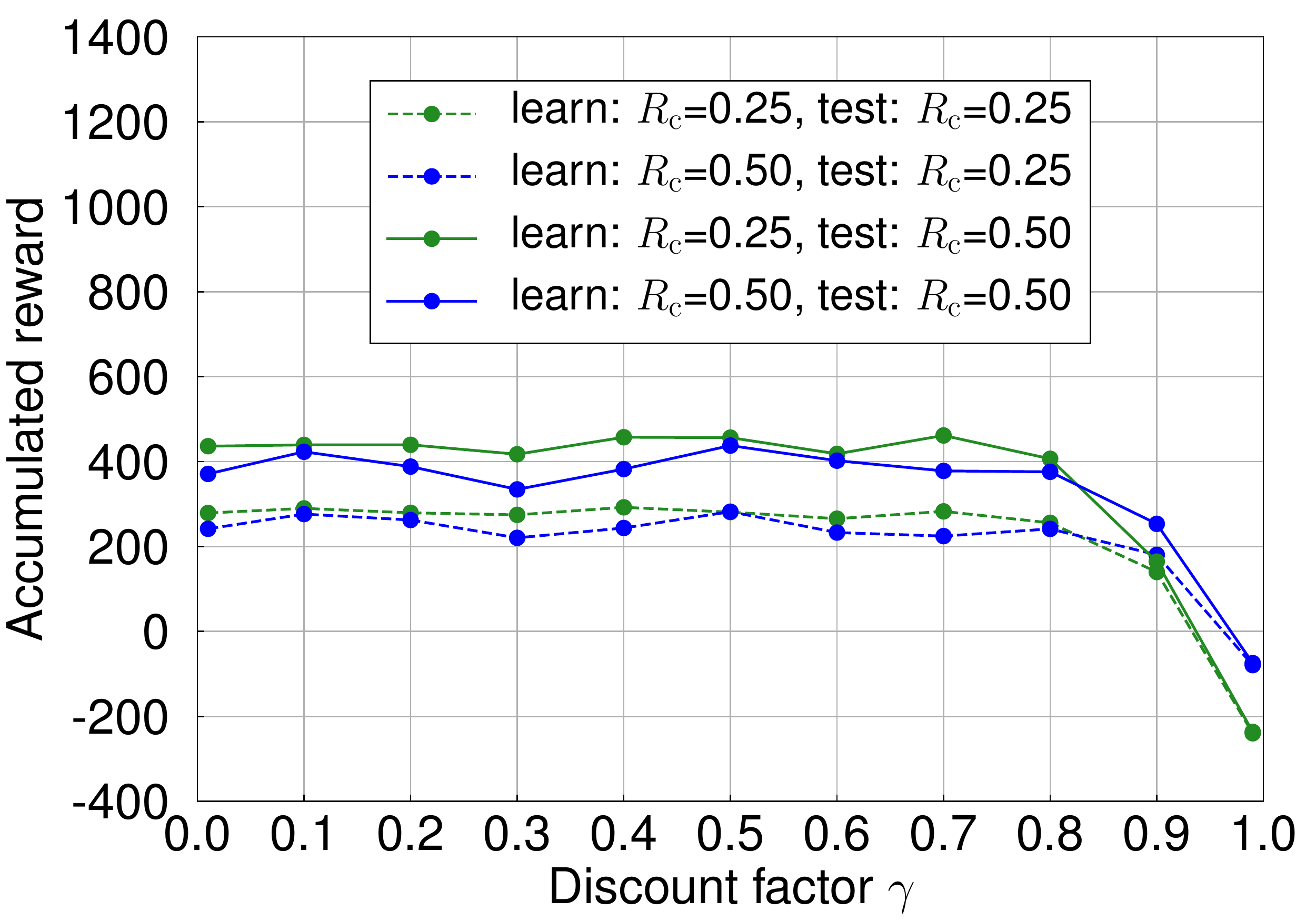}
  \caption{Accumulated reward for different discount factors $\gamma$
  when $\lambda=0.02$\,veh/m/lane, $\rmm=0.4$, and \PTCL is used as the state type.
  Uncontrollable vehicles move at constant velocities.}
  \label{fig:discountrwd}
\end{figure}

\begin{figure}[t] \centering
  \includegraphics[width=\figsizeData]{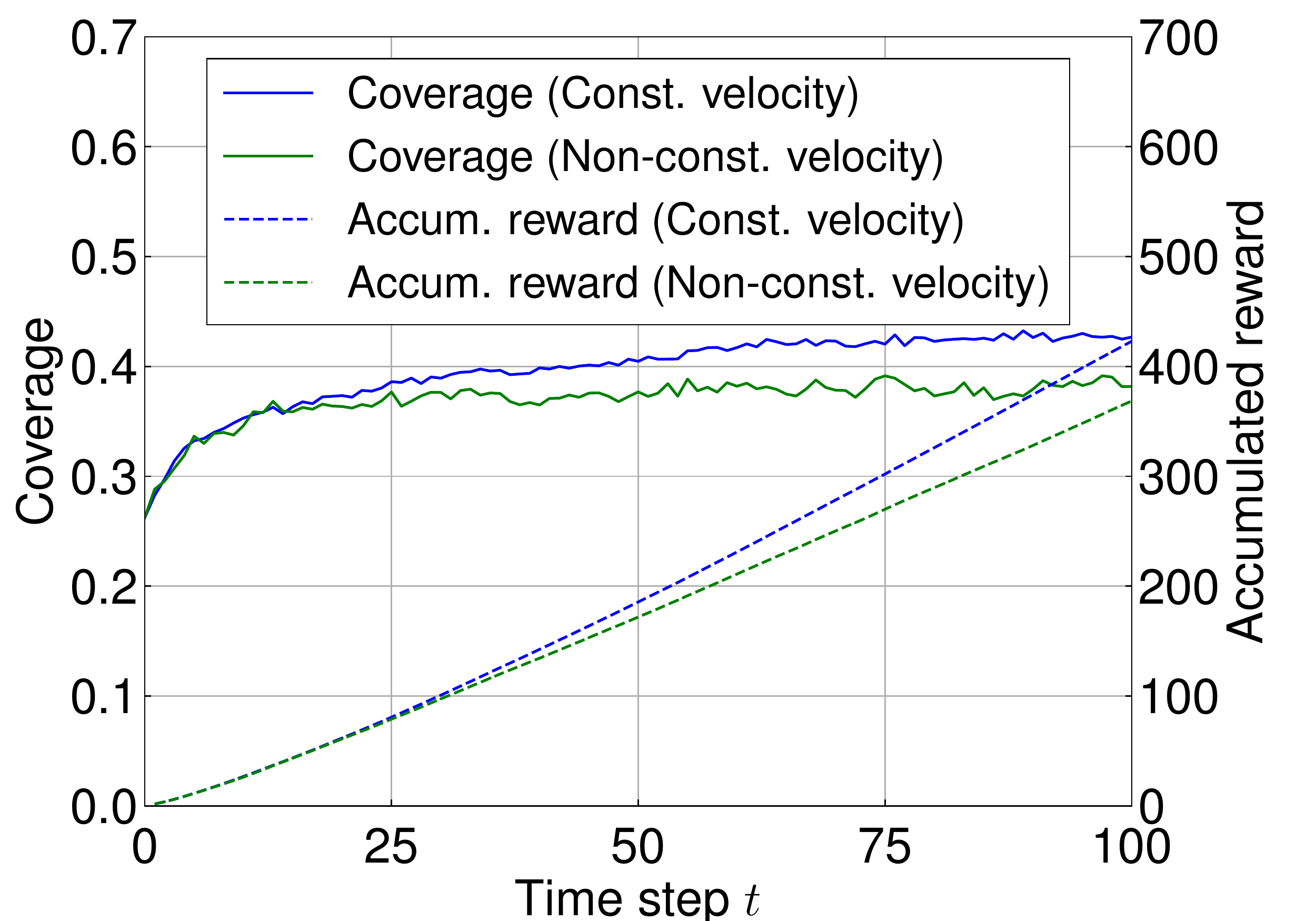}
  \caption{Coverage and accumulated reward as functions of time steps
  when $\lambda=0.02$\,veh/m/lane, $\rmm=0.4$, $\rctrl=0.5$, and \PTCL is used as the state type.
  Accumulated rewards increase linearly even after coverage becomes constant.}
  \label{fig:coverage_step}
\end{figure}

\begin{figure}[t] \centering
  \includegraphics[width=\figsizeData]{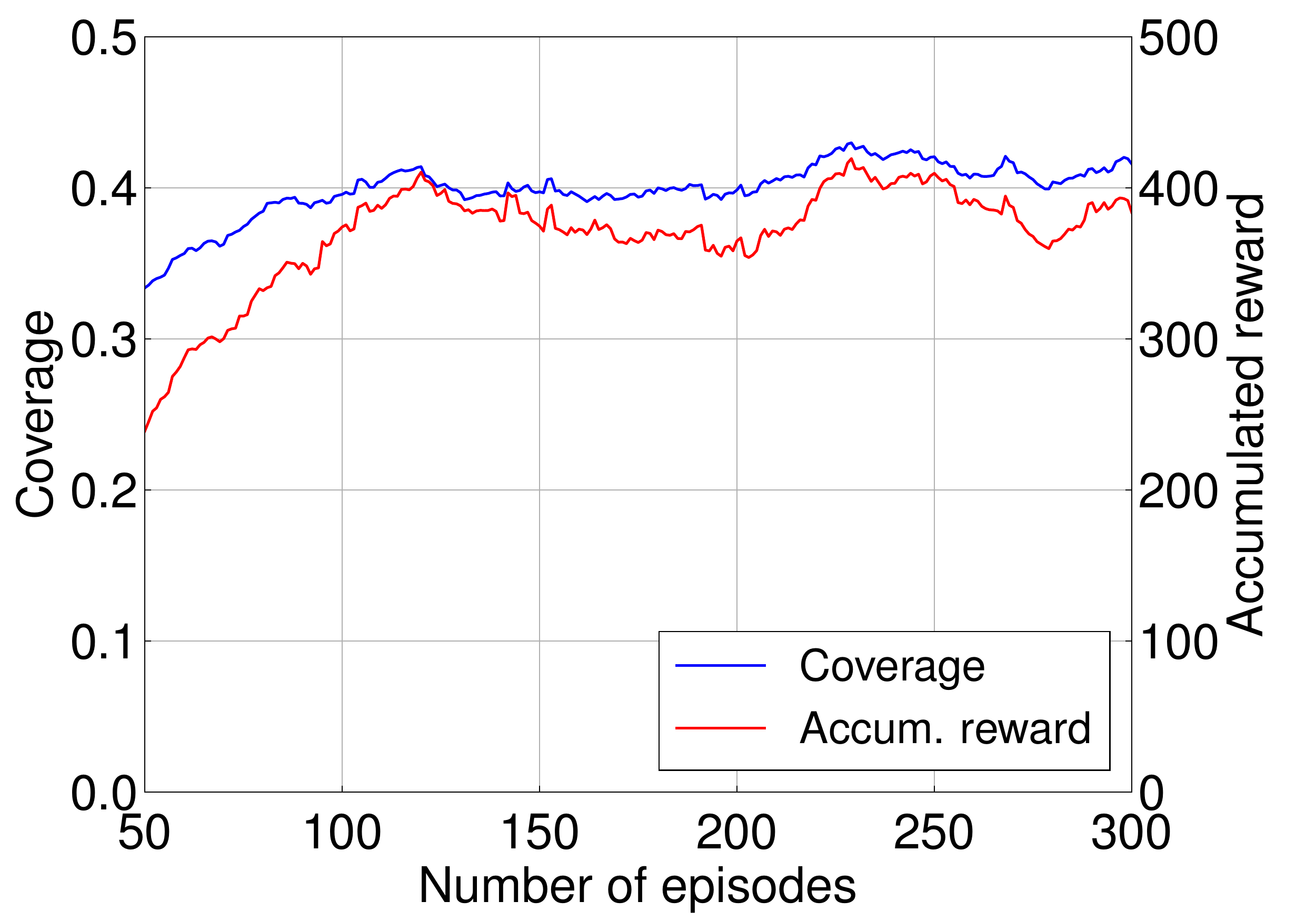}
  \caption{Moving average of coverage and accumulated reward
  as functions of the number of training episodes
  when $\lambda=0.02$\,veh/m/lane, $\rmm=0.4$, $\rctrl=0.5$, and \PTCL is used as the state type.
  Uncontrollable vehicles move at constant velocities.}
  \label{fig:moving_avg}
\end{figure}

\begin{figure}[t] \centering
  \includegraphics[width=\figsizeData]{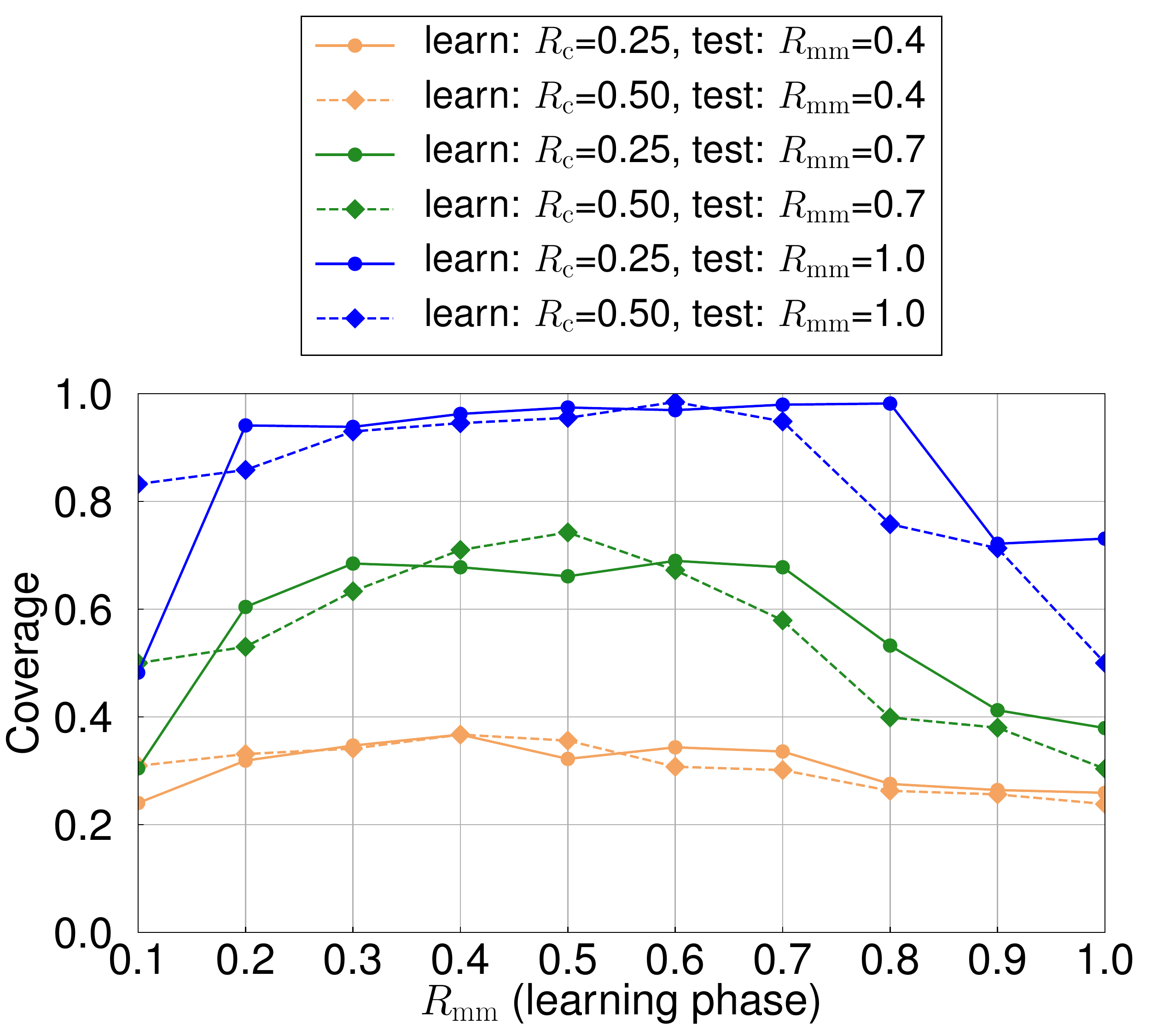}
  \caption{Coverage of models learned in different learning conditions $\rmm$ and $\rctrl$
  when $\lambda=0.02$\,veh/m/lane.
  \PT is used as the state type and uncontrollable vehicles move at constant velocities.
  The models learned when $0.3 \le \rmm \le 0.7$ show relatively higher performance.}
  \label{fig:comp_model}
\end{figure}

\subsection{Performance of Reinforcement-Learning-Based Method}
The coverage performance with three types of states
\PT, \PTCL, and \PTDL when $\lambda=0.02$\,veh/m/lane and $\rmm=0.4$
is shown in Fig.~\ref{fig:rlstate}.
It can be seen that
\PTCL and \PTDL achieve higher performance than \PT,
which means that additional relay length information
improves the performance of RL-based methods.

Figures~\ref{fig:discount} and \ref{fig:discountrwd} show
the coverage performance and accumulated reward with different discount factors
when $\lambda=0.02$\,veh/m/lane, $\rmm=0.4$,
and the state type is \PTCL.
When the discount factor $\gamma$ is greater than 0.8, the coverage decreases.
In particular, when $\gamma=0.99$,
the accumulated rewards are less than zero,
which means that agents failed to learn a reasonable policy
under the hyper parameter settings and the simulation scenarios.

The coverage and accumulated reward as functions of time step $t$
are shown in Fig.~\ref{fig:coverage_step}.
Simulations are performed under two conditions
where non-mmWave vehicles and uncontrollable vehicles run at equal and constant velocities
and where they change their velocities randomly.
In both conditions, the coverage increases as the agents change their positions.
Coverage is greater
when uncontrollable vehicles move at the same velocity,
than when uncontrollable vehicles move at different velocities.
Accumulated reward increases linearly even when $t>75$
because nearly constant rewards are given to agents after the coverage converges.

Coverage and accumulated reward as functions of the number of episodes
in learning phases are shown in Fig.~\ref{fig:moving_avg}.
50-episode moving averages are shown
because the achievable coverage in each episode varies
due to the random variables such as the number of vehicles and initial vehicle positions.
It is shown that coverage and accumulated reward increase at first,
and then do not increase when the number of episodes is greater than 120.
This is because
various experiences improve agents' strategies at first,
whereas additional experiences do not contribute to improving the strategies as much
after the agents have learned from many experiences.

In Fig.~\ref{fig:comp_model}, the coverage performance with different models
obtained in different learning conditions is shown.
We trained 20 models in learning environments with varying $\rctrl$ and $\rmm$,
transferred them to test environments with $\rmm=0.4,0.7,1.0$,
and evaluated coverage of the position controls using the models. 
The models learned when $0.3\le\rmm\le 0.7$ show approximately the same performance,
which means these models can be applied
when the penetration ratio is changed from that in the learning phases. 
On the other hand, the models learned when $\rmm \ge 0.9$
show lower performance.
When there are many mmWave vehicles in the learning phases,
large coverage can be achieved without movement of agents.
Therefore, the agents cannot learn aggressive moving policies.
This is why the models learned with large $\rmm$ show lower performance.
The models learned when $\rctrl=0.25$ and $\rmm=0.1$ also show lower performance.
This is because the number of agents is too small to learn policies that increase coverage.

\begin{figure}[t] \centering
  \includegraphics[width=\figsizeData]{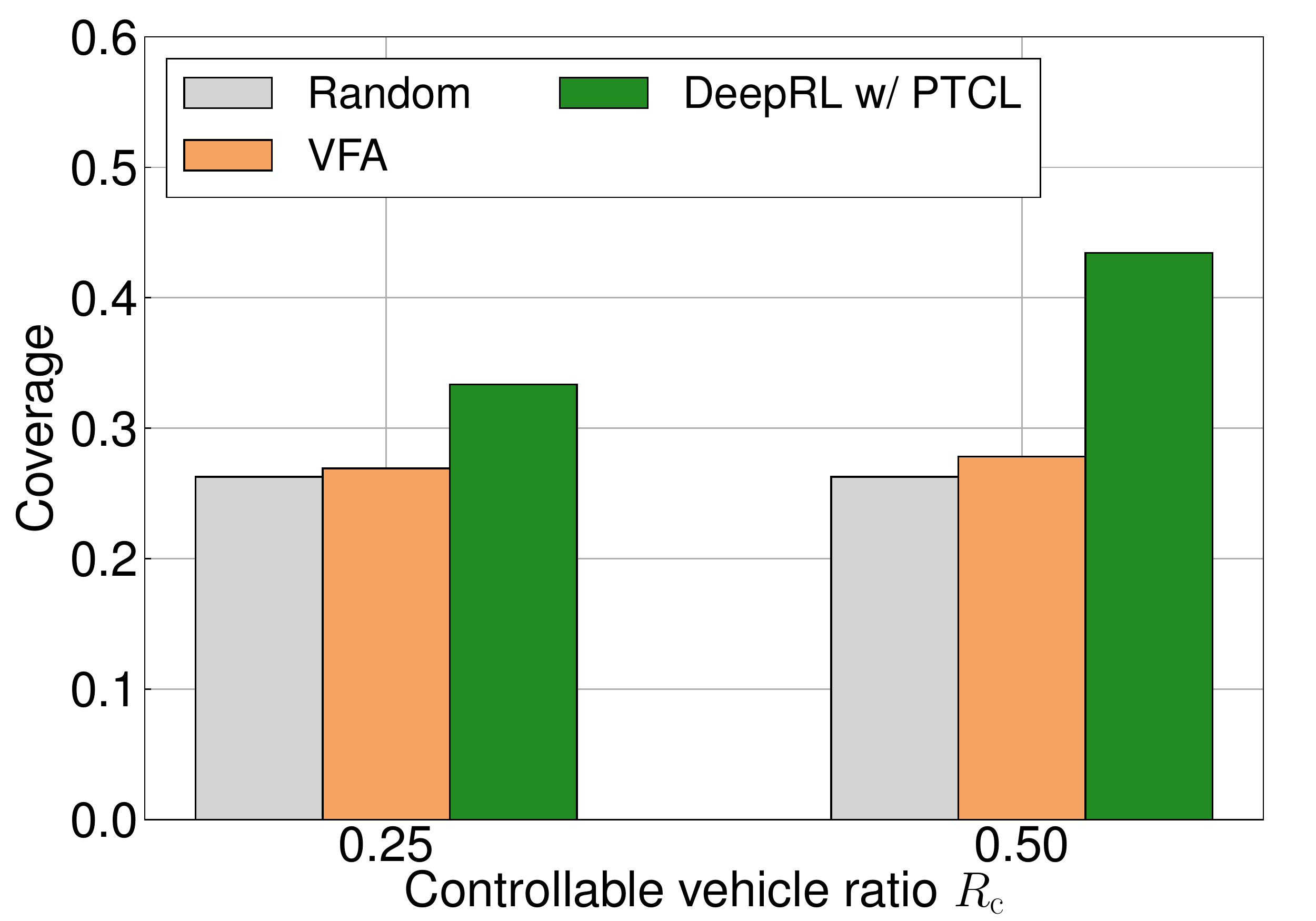}
  \caption{Coverage of different movable methods
  when $\lambda=0.02$\,veh/m/lane, $\rmm=0.4$ and uncontrollable vehicles move at constant velocities.
  RL shows higher performance than random and VFA.}
  \label{fig:type_rmm40}
\end{figure}

\begin{figure}[t] \centering
  \includegraphics[width=\figsizeData]{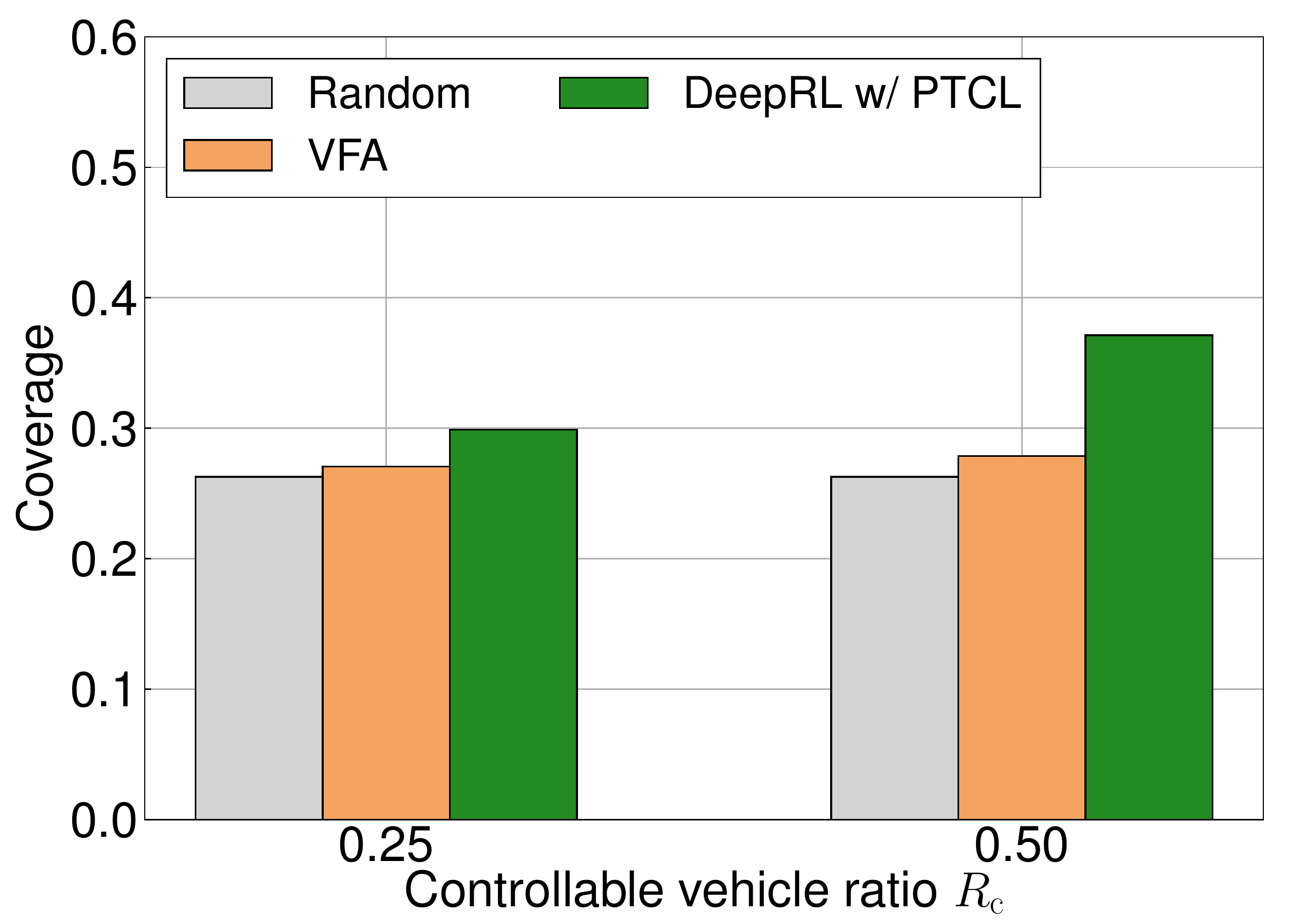}
  \caption{Coverage of different movable methods
  when $\lambda=0.02$\,veh/m/lane, $\rmm=0.4$ and uncontrollable vehicles move at random velocities.
  RL shows higher performance than random and VFA.}
  \label{fig:type_rmm40_rand}
\end{figure}

\begin{figure}[t] \centering
  \includegraphics[width=\figsizeData]{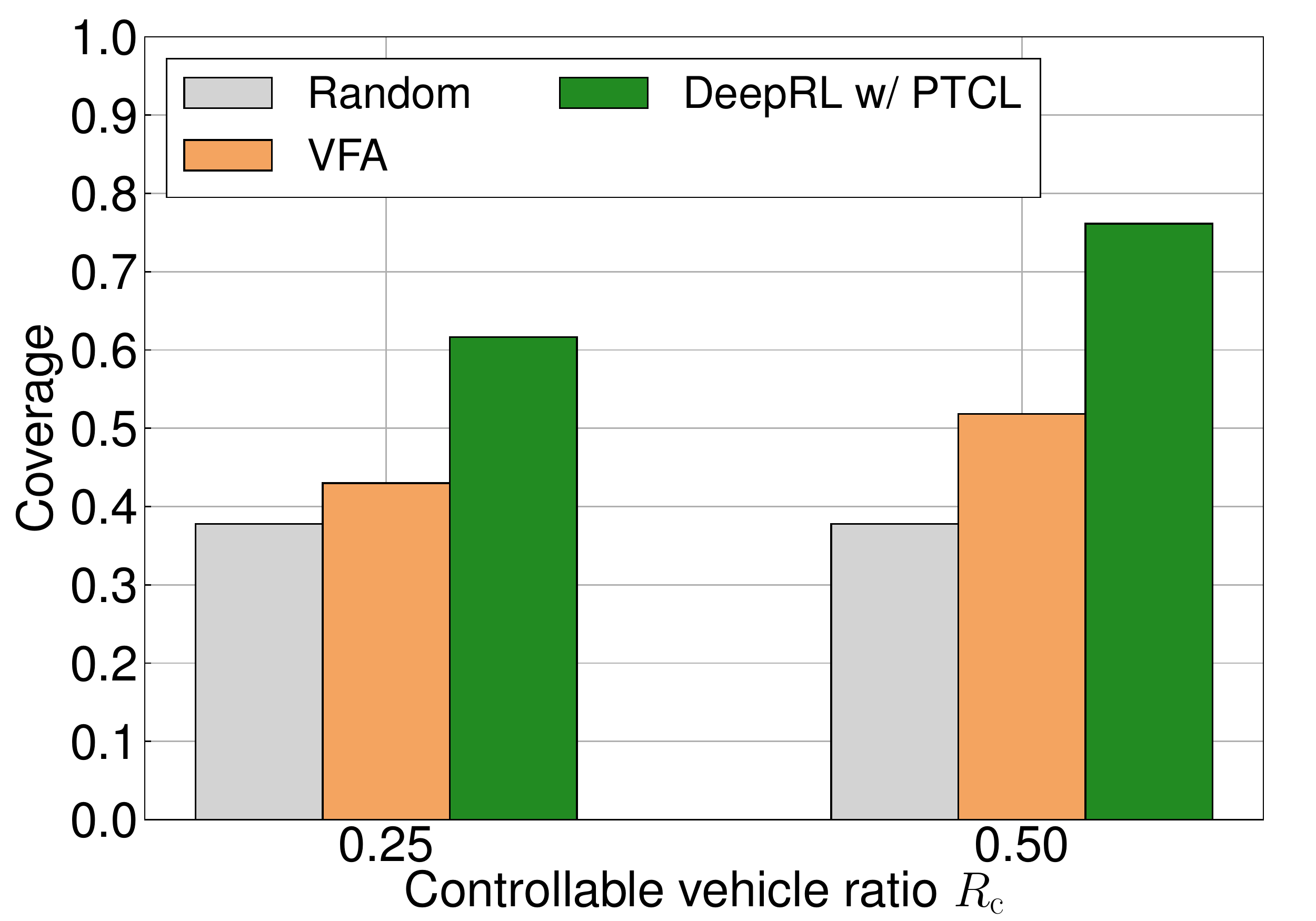}
  \caption{Coverage of different movable methods
  when $\lambda=0.01$\,veh/m/lane, $\rmm=1$ and uncontrollable vehicles move at constant velocities.}
  \label{fig:type_rmm100}
\end{figure}

The coverage performance of the RL-based method with \PTCL
is compared with that of random positions and VFA
under conditions
when $\lambda=0.02$\,veh/m/lane, $\rmm=0.4$,
and uncontrollable vehicles move at constant and equal velocities
in Fig.~\ref{fig:type_rmm40}.
As a comparative method,
we use exponential VFA \cite{NovelDeploymentVF}
with parameters
$W_a=1$, $W_r=10000$,
$\beta_1=\beta_2=2$,
and $D_\mathrm{th}=50$\,m.
Details of these parameters are described in \cite{NovelDeploymentVF}.
The coverage achieved with the RL-based method is higher
than the coverage of random positions and VFAs.
For example, the coverage obtained from the RL-based method
is about 1.7 times the coverage obtained with random positions when $\rctrl=0.5$.
Although vehicles with VFA adjust distances between themselves
in order to connect with each other,
they do not consider blockage of mmWave communications.
On the other hand,
vehicles with RL agents
optimize their actions to expand their coverage
following the learned policy models.
Consequently, the performance of DeepRL exceeds that of VFA.

The coverage performance with different method
is compared with that of random positions and VFA
under conditions
when $\lambda=0.02$\,veh/m/lane, $\rmm=0.4$,
and uncontrollable vehicles move at different velocities
in Fig.~\ref{fig:type_rmm40_rand}.
It is notable that the performance of the proposed method
exceeds that of VFA even when the vehicles move at different velocities.

Coverage performance
when $\lambda=0.01$\,veh/m/lane, $\rmm=1$,
and uncontrollable vehicles move at constant and equal velocities
is shown in Fig.~\ref{fig:type_rmm100}.
The training of the RL-based methods were conducted
under the conditions with $\lambda=0.02$ and $\rmm=0.4$,
and the model transferred to this evaluation environment.
The RL performance shows larger coverages than the results with $\rmm=0.4$ in Fig.~\ref{fig:type_rmm40}
because there are fewer obstacles that disrupt the movement of the controllable vehicles (i.e., RL agents)
and thus, they easily move to positions where long relays can be achieved.
Furthermore, VFAs show lower performance than the RL-based methods
because they rely on the assumption that
all vehicles are controllable
while they are not in the environment.

\subsection{Comparisons with Other RL Algorithm}
Here, we discuss other RL algorithms
compared with one used in the proposed method for the position control problem.
As introduced in \cite{nguyen2017system}, there are many RL algorithms.
The table-based Q-learning is one of most widely used algorithm for various problems.
However, the algorithm is not suitable for our problem
because it requires large memory capacity,
whereas it is difficult for vehicles to have large memory.
In the table-based Q-learning,
an action-value function is represented as a table, called Q table,
and the memory capacity required to store the table
is proportional to the number of values that the state $\si_{i,t}$ can take.
In our problem, it increases exponentially with the observation range $X\times Y$;
thus, the table-based Q-learning is difficult to apply for our problem
when the observation range becomes large.

DQN \cite{dqn} can be a candidate algorithm for our problem.
DQN utilizes DNN to approximate an action-value function
instead of using the Q table;
thus, the required memory capacity can be small when the state is large.
However,
DQN cannot leverage experiences obtained by multiple vehicles
while the vehicles try to learn similar strategies,
because the algorithm is designed for single-agent problems.

There are several multi-process DeepRL algorithms,
which can be applied to multi-agent environments.
The DeepRL algorithms can leverage experiences of multiple agent to update a model;
thus, it can learn a strategy faster
and the strategy can be achieved higher performance
than single-agent RL algorithm \cite{a3c,gorila}.
We applied A3C \cite{a3c} to our problems in this paper
as a state-of-the-art multi-process DeepRL algorithm,
and we showed that the proposed method increases coverage in our problem.
Since the focus of our paper
is not on designing or determining the best RL algorithm for the problem,
the detailed performance comparison between A3C and other multi-agent RL algorithms
is out of our scope.

\section{Conclusions and Future Works} \label{sec:conclusion}
We proposed a vehicle position control method for mmWave vehicular networks
that increases coverage through multi-hop V2V relaying.
We adopted A3C,
which is one of the state-of-the-art of RL methods,
to obtain practical movement strategies
and designed states with relay length information
that improved system performance in terms of increasing coverage.
We also showed that the RL-based strategy
achieved 1.7 times the coverage of random positions
when the penetration ratio of mmWave vehicles was 40\%
and half of the mmWave vehicles employed the proposed RL algorithm.
Moreover, the RL-based strategy increased coverage,
even when the vehicle density and penetration ratio
of mmWave vehicles differed from those of the learning phase.

Our future work includes evaluating the performance of the proposed method
with metrics other than relay length.
One of the advantages of RL is applicability
for tasks with other objectives by redesigning the reward.
For example,
the reliability and connectivity of the multi-hop network
is more important than relay length
when the vehicular network is used for safety applications
or vehicular cloud computing.
The proposed method has potential to achieve such objectives;
thus, we will demonstrate the performance in our future work. 

A method to share and aggregate models learned by the vehicles
to improve their performance
is also included in our future work.
Current work assumed that the model sharing is conducted ideally,
i.e., the sharing is done without delay, loss, and bandwidth consumption.
The model-sharing method should be designed
in consideration of scalability, delay, loss,
and bandwidth consumption of wireless communications.

\section*{Acknowledgment}
This work was supported in part by JSPS KAKENHI Grant Number 17H03266,
KDDI Foundation,
and Tateisi Science and Technology Foundation.

\bibliographystyle{ieicetr}% bib style
\bibliography{main}% your bib database
% \begin{thebibliography}{99}% more than 9 --> 99 / less than 10 --> 9
% \bibitem{}
% \end{thebibliography}

%\profile{}{}
%\profile*{}{}% without picture of author's face
% \profile{Akihito Taya}{
\profile[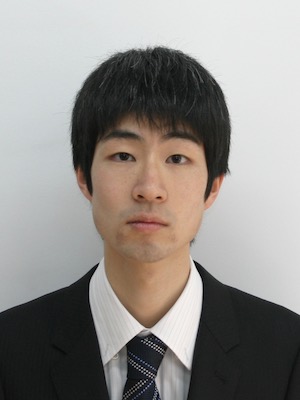]{Akihito Taya}{
received the B.E.\ degree in Electrical and Electronic Engineering from Kyoto University in 2011. 
He received the master degree in Communications and Computer Engineering, Graduate School of Informatics from Kyoto University, Kyoto, Japan, in 2013.
He joined Hitachi, Ltd. in 2013, where he perticipated in the development of computer clusters.
He is currently working toward a Ph.D.\ degree at the Graduate School of Informatics, Kyoto University.
His current research interests include vehicular communications and applications of machine learning.
He is a student member of the IEEE, ACM, and IEICE.
}

\pagebreak

% \profile{Takayuki Nishio}{
\profile[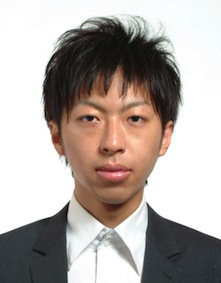]{Takayuki Nishio}{
received the B.E.\ degree in Electrical and Electronic Engineering from Kyoto University in 2010. 
He received the master and Ph.D. degrees in Communications and Computer Engineering, Graduate School of Informatics from Kyoto University, Kyoto, Japan, in 2012 and 2013, respectively.
From 2012 to 2013, he was a research fellow (DC1) of the Japan Society for the Promotion of Science (JSPS).
Since 2013, He is an Assistant Professor in Communications and Computer Engineering, Graduate School of Informatics, Kyoto University. From 2016 to 2017, he was a visiting researcher in Wireless Information Network Laboratory (WINLAB), Rutgers University, United States. 
His current research interests include mmWave networks, wireless local area networks, application of machine learning, and sensor fusion in wireless communications. He received IEEE Kansai Section Student Award in 2011, the Young Researcher's Award from the IEICE of Japan in 2016, and Funai Information Technology Award for Young Researchers in 2016. He is a member of the IEEE, ACM, IEICE.
}

% \pagebreak

% \profile{Masahiro Morikura}{
\profile[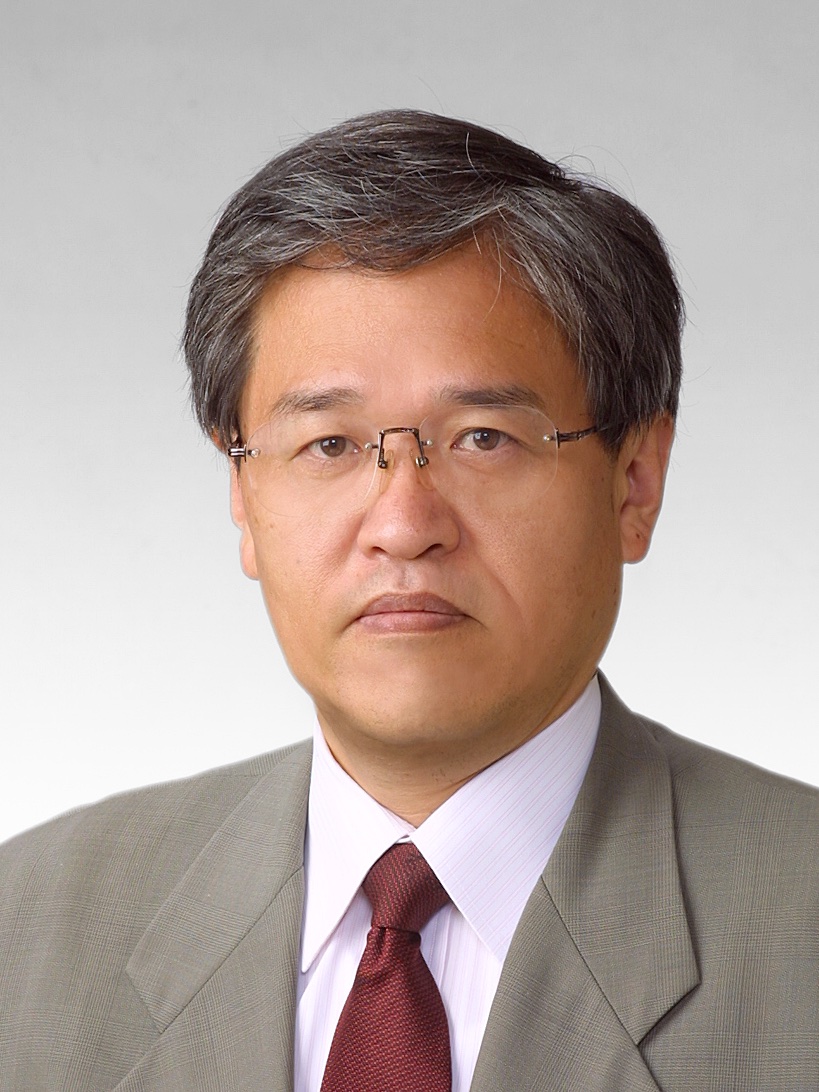]{Masahiro Morikura}{
% received his B.E., M.E., and Ph.D.\ degrees in electronics engineering from Kyoto University, Kyoto, Japan in 1979, 1981 and 1991, respectively. He joined NTT in 1981, where he was engaged in the research and development of TDMA equipment for satellite communications. From 1988 to 1989, he was with the Communications Research Centre, Canada, as a guest scientist. From 1997 to 2002, he was active in the standardization of the IEEE 802.11a based wireless LAN. His current research interests include WLANs and M2M wireless systems. He received the Paper Award and the Achievement Award from IEICE in 2000 and 2006, respectively. He also received the Education, Culture, Sports, Science and Technology Minister Award in 2007 and Maejima Award in 2008. Dr. Morikura is now a professor in the Graduate School of Informatics, Kyoto University. He is a member of the IEEE.
Masahiro Morikura received his B.E., M.E., and Ph.D.\ degrees in electronics engineering from Kyoto University, Kyoto, Japan in 1979, 1981 and 1991, respectively.
He joined NTT in 1981, where he was engaged in the research and development of TDMA equipment for satellite communications. From 1988 to 1989,
he was with the Communications Research Centre, Canada, as a guest scientist.
From 1997 to 2002, he was active in the standardization of the IEEE 802.11a based wireless LAN.
His current research interests include WLANs and M2M wireless systems.
He received the Paper Award and the Achievement Award from IEICE in 2000 and 2006, respectively.
He also received the Education, Culture, Sports, Science and Technology Minister Award in 2007 and Maejima Award in 2008, and the Medal of Honor with Purple Ribbon from Japan's Cabinet Office in 2015.
Dr. Morikura is now a professor in the Graduate School of Informatics, Kyoto University. He is a member of the IEEE.
}

% \pagebreak

% \profile{Koji Yamamoto}{received the B.E.\ degree in electrical and electronic engineering from Kyoto University in 2002, and the M.E.\ and Ph.D.\ degrees in Informatics from Kyoto University in 2004 and 2005, respectively.
\profile[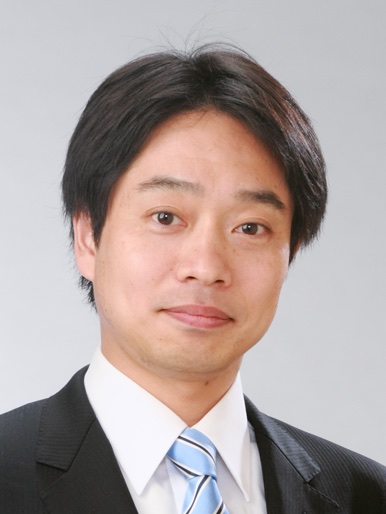]{Koji Yamamoto}{received the B.E.\ degree in electrical and electronic engineering from Kyoto University in 2002, and the M.E.\ and Ph.D.\ degrees in Informatics from Kyoto University in 2004 and 2005, respectively.
From 2004 to 2005, he was a research fellow of the Japan Society for the Promotion of Science (JSPS).
Since 2005, he has been with the Graduate School of Informatics, Kyoto University, where he is currently an associate professor.
From 2008 to 2009, he was a visiting researcher at Wireless@KTH, Royal Institute of Technology (KTH) in Sweden.
He serves as an editor of IEEE Wireless Communications Letters from 2017 and the Track Co-Chairs of APCC 2017 and CCNC 2018.
His research interests include radio resource management and applications of game theory.
He received the PIMRC 2004 Best Student Paper Award in 2004, the Ericsson Young Scientist Award in 2006.
He also received the Young Researcher's Award, the Paper Award, SUEMATSU-Yasuharu Award from the IEICE of Japan in 2008, 2011, and 2016, respectively, and IEEE Kansai Section GOLD Award in 2012.
He is a member of the IEEE.
}

\end{document}